\documentclass[12pt, letterpaper]{article}

\usepackage{amssymb}
\usepackage{amsmath}
\usepackage{verbatim}

\usepackage[nosort]{cite}
\usepackage[hyperref,bulletsep]{collect}
\usepackage{hyperref}

\mathversion{normal}

\makeatletter \@addtoreset{equation}{section} \makeatother

\newcommand{\gen}[1]{\mathfrak{#1}}
\newcommand{\alg}[1]{\mathfrak{#1}}

\ifx\genfrac\sdflkaj

\else

\fi
\newcommand{\sfrac}[2]{{\textstyle\frac{#1}{#2}}}
\ifx\half\sdflkaj
\newcommand{\half}{\sfrac{1}{2}}
\fi

\newcommand{\inner}[1]{\langle #1 \rangle}
\newcommand{\comm}[2]{[#1,#2]}
\newcommand{\acomm}[2]{\{#1,#2\}}

\newcommand{\gcomm}[2]{[#1,#2\}}
\newcommand{\state}[1]{\mathopen{\big|}#1\mathclose{\bigr \rangle}}

\let\oldPhi=\Phi
\let\oldPsi=\Psi
\renewcommand{\Phi}{\mathnormal{\oldPhi}}
\renewcommand{\Psi}{\mathnormal{\oldPsi}}

\newcommand{\nln}{\nonumber\\}
\newcommand{\nl}{\nonumber\\&&\mathord{}}
\newcommand{\earel}[1]{\mathrel{}&#1&\mathrel{}}
\newcommand{\eq}{\earel{=}}
\newenvironment{myeqnarray}{\arraycolsep0pt\begin{eqnarray}}{\end{eqnarray}\ignorespacesafterend}
\newenvironment{myeqnarray*}{\arraycolsep0pt\begin{eqnarray*}}{\end{eqnarray*}\ignorespacesafterend}

\def\[{\begin{equation}}
\def\]{\end{equation}}
\def\<{\begin{myeqnarray}}
\def\>{\end{myeqnarray}}

\ifx\href\asklfhas\newcommand{\href}[2]{#2}\fi
\newcommand{\arxivno}[1]{\href{http://arxiv.org/abs/#1}{#1}}

\begin{document}
\thispagestyle{empty}
\begin{flushright}\footnotesize
\texttt{\arxivno{arXiv:1111.0083}}\\ 
\texttt{CALT-68-2852}
\end{flushright}
\vspace{.5cm}

\begin{center}%
{\Large\textbf{\mathversion{bold}%
From Scattering Amplitudes  \\  to the Dilatation Generator  in $\mathcal{N}=4$  SYM}\par}\vspace{1.5cm}%

\textsc{Benjamin~I.~Zwiebel} \vspace{8mm}

\textit{ California Institute of Technology \\ 
Pasadena, CA 91125, USA}%
\vspace{4mm}

\texttt{bzwiebel@caltech.edu}\par\vspace{1.5cm}

\textbf{Abstract}\vspace{7mm}

\begin{minipage}{14.7cm}
The complete spin chain representation of the planar $\mathcal{N}=4$ SYM dilatation generator has long been known at one loop, where it involves leading nearest-neighbor $2 \rightarrow 2$ interactions. In this work we use superconformal symmetry to derive the unique solution for the leading $L \rightarrow 2$ interactions of the planar dilatation generator for arbitrarily large $L$. We then propose that these interactions are given by the scattering operator that has $\mathcal{N}=4$ SYM tree-level scattering amplitudes as matrix elements.  We provide compelling evidence for this proposal, including explicit checks for $L=2,3$ and a proof of consistency with superconformal symmetry. 
\end{minipage}

\end{center}

\newpage
\setcounter{page}{1}
\renewcommand{\thefootnote}{\arabic{footnote}}
\setcounter{footnote}{0}


\section{Introduction}

Computations of gauge theory correlation functions or scattering amplitudes typically become intractable beyond low orders in perturbation theory. For $\mathcal{N}=4$ planar supersymmetric Yang-Mills theory, dualities and integrability improve the situation tremendously. The spectrum of anomalous dimensions of $\mathcal{N}=4$ SYM, or equivalently the energy spectrum of strings in the dual $AdS_5 \times S^5$ string theory, are given by solutions of a system of Bethe ansatz equations \cite{Minahan:2002ve,Arutyunov:2004vx,Staudacher:2004tk,Beisert:2005fw,Beisert:2006ez}, with wrapping corrections incorporated through thermodynamic Bethe ansatz or Y-system equations \cite{Gromov:2009tv,Bombardelli:2009ns,Arutyunov:2009ur,Gromov:2011cx}.  See \cite{Beisert:2010jr} for a recent review. 

Likewise, integrability has played a central role in progress in computing on-shell scattering amplitudes of $\mathcal{N}=4$ SYM. This progress has also revealed multiple new dualities.  Following Witten's proposal of a  twistor string dual \cite{Witten:2003nn}, the BCFW recursion relations for all tree-level gluon amplitudes were found \cite{Britto:2004ap,Britto:2005fq}, and later extended to all tree-level amplitudes \cite{Brandhuber:2008pf,ArkaniHamed:2008gz}. Moreover, the calculation of MHV scattering amplitudes  is dual to a Wilson loop calculation both at strong coupling \cite{Alday:2007hr} and at weak coupling \cite{Drummond:2007aua, Brandhuber:2007yx, Drummond:2007cf}. This implies that amplitudes have a dual conformal symmetry, which extends  to dual superconformal symmetry \cite{Drummond:2008vq} that originates in a fermionic T-duality symmetry of the $AdS_5 \times S^5$ string theory \cite{Berkovits:2008ic,Beisert:2008iq}. Confirming integrability, the ordinary and dual superconformal symmetry generate the Yangian of $\alg{psu}(2,2|4)$, which has been shown directly for tree-level  \cite{Drummond:2009fd}  and one-loop \cite{Sever:2009aa,Beisert:2010gn} amplitudes\footnote{Integrability also enables a Y-system calculation of MHV amplitudes at strong coupling \cite{Alday:2009dv,Alday:2010vh}.}.
In fact, Grassmannian duality \cite{ArkaniHamed:2009dn} ensures the Yangian symmetry for the leading singularities of scattering amplitudes \cite{ArkaniHamed:2009vw}. This Grassmannian duality and the Yangian symmetry extend to the generalization of BCFW recursion relations to the all-loop integrand for scattering amplitudes \cite{ArkaniHamed:2010kv}. 

 An important motivation for focusing on  planar $\mathcal{N}=4$ SYM  is the possibility that its simplifications due to dualities and integrability will help reveal important insights or new technical methods that can be applied to realistic gauge theories, such as QCD. An inspiring example is that generalizations of the BCFW recursion relations are now an important calculation method for the LHC \cite{Ita:2011hi}.    

Seeking to transfer additional lessons from $\mathcal{N}=4$ SYM, it makes sense to try to complete our understanding of this gauge theory's special properties. Notably, there is no proof of integrability for the spectral problem at weak coupling beyond one loop. In fact,  the dilatation generator, whose eigenvalues give the spectrum of anomalous dimensions, is only known at one loop  \cite{Lipatov:1997vu,Dolan:2001tt,Beisert:2003jj} and at higher loops in special sectors \cite{Beisert:2003ys,Belitsky:2005bu,Zwiebel:2008gr,Bargheer:2009xy}.  In the planar limit,  we represent local operators as spin chain states, and then the dilatation generator acts as a spin chain Hamiltonian. This spin chain Hamiltonian has the unusual property of including multisite length-changing  interactions \cite{Beisert:2003ys}. These interactions replace $L$ initial sites  with $L'$ final sites, and we label the $L \rightarrow L'$ contributions to the dilatation generator  $\gen{D}_{L \rightarrow L'}$. As explained below in Section \ref{sec: beyondleadingorder}, we use a nonstandard normalization, which treats $L$ and $L'$ differently. In this normalization, $L \rightarrow L'$ interactions first appear at $\mathcal{O}(g^{2 L' -2})$, where $g$ is related to the `t Hooft coupling as $g^2 = \lambda/(16 \pi^2)$. We will restrict our attention to these leading interactions  ($\mathcal{O}(g^{2 L' -2})$) in this work. Note that in this normalization and regardless of the number of initial sites, leading interactions with a single final site are $\mathcal{O}(g^0)$, and leading interactions with two final sites are $\mathcal{O}(g^2)$.

The exact superconformal $\alg{psu}(2,2|4)$ symmetry of $\mathcal{N}=4$ SYM tightly constrains the dilatation generator. In particular, the anomalous part of the dilatation generator commutes with all other  symmetry generators. In the classical ``linear'' approximation,  these generators only have $1 \rightarrow 1$ interactions. However, they also receive corrections involving multisite length-changing interactions. In this work we restrict to the leading order in $g$, which is $\mathcal{O}(g^0)$ for the superconformal generators (interactions with one final site) and $\mathcal{O}(g^2)$ for the (anomalous) dilatation generator (interactions with two final sites).  Then, only the superconformal generators $\gen{S}$, $\dot{\gen{S}}$, and $\gen{K}$ have corrections, which are $2 \rightarrow 1$ interactions\footnote{$\gen{K}$ has  $3 \rightarrow 1$ interactions too.}. These corrections were derived in \cite{Zwiebel:2007th}. For the dilatation generator, at $\mathcal{O}(g^2)$, leading interactions $\gen{D}_{L \rightarrow 2}$ for all $L \geq 2$ appear. 

We will show that superconformal symmetry fixes $\gen{D}_{L \rightarrow 2}$ uniquely. We will derive a few equations that give $\gen{D}_{L \rightarrow 2}$ in terms of $\gen{D}_{(L-1) \rightarrow 2}$, $\gen{D}_{(L-2) \rightarrow 2}$ and the (known) corrections to $\gen{K}$. Since $\gen{D}_{2 \rightarrow 2}$ is the known one-loop dilatation generator, this uniquely defines all $\gen{D}_{L \rightarrow 2}$. Recall that the one-loop dilatation generator is fixed by superconformal symmetry \cite{Beisert:2004ry}. We see that superconformal symmetry is just as powerful for all of the leading dilatation generator interactions with two final sites. In the standard normalization, the $\gen{D}_{L \rightarrow 2}$ interactions we find would be $\mathcal{O}(g^L)$, or ``$L/2$-loop'' interactions.

Because the dilatation generator is Hermitian, $\gen{D}_{2 \rightarrow L}$ follow from Hermitian conjugation of  $\gen{D}_{L \rightarrow 2}$. Of course, we still need the $\gen{D}_{L \rightarrow L'}$ for both $L,L' > 2$ to have the full ``leading'' dilatation generator that corresponds to the asymptotic Bethe ansatz. Our derivation of  $\gen{D}_{L \rightarrow 2}$ depends significantly on the restricted possibilities for interactions with two final sites. A continued direct approach seems unlikely to be sufficient to find the remaining unknown $\gen{D}_{L \rightarrow L'}$, which are less constrained. As described above,  the progress on $\mathcal{N}=4$ SYM frequently uses dual descriptions.  As a step in that direction for the dilatation generator,  in this work we will precisely relate  the $\gen{D}_{L \rightarrow 2}$ to tree-level scattering amplitudes in $\mathcal{N}=4$ SYM\footnote{At first glance,  it seems surprising to relate the leading \emph{$(L/2)$-loop} dilatation generator to \emph{tree-level} scattering amplitudes. However, the powers of $g$ for these dilatation generator components arise from vertices connecting spacetime points in Feynman graphs, and not from momentum loops.  The contributing graphs for the leading $\gen{D}_{L \rightarrow 2}$ do not have momentum loops, as is the case for tree-level scattering amplitudes.}. 

Hints of such a connection  between the dilatation generator and scattering amplitudes can be seen in work of Bargheer, Beisert,  Galleas, Loebbert and  McLoughlin \cite{Bargheer:2009qu}, which we now review briefly.  Tree-level scattering amplitudes in $\mathcal{N}=4$ SYM are annihilated by superconformal generators for generic momenta, but not for configurations with collinear momenta.  This anomaly occurs because a single massless particle is indistinguishable from multiple collinear massless particles carrying the same total momentum.  \cite{Bargheer:2009qu} showed that one can deform the anomalous superconformal generators ($\gen{S}$, $\bar{\gen{S}}$, $\gen{K}$) such that the deformed generators do annihilate the amplitude generating functional. To account for the above-mentioned indistinguishability, the deformations replace one external particle with two (or three) particles.   As observed in  \cite{Bargheer:2009qu}, this particle-number-changing deformation is analogous to the length-changing corrections to the spin chain generators of the spectral problem.

We note that at $\mathcal{O}(g^0)$ the same generators  ($\gen{S}$, $\dot{\gen{S}}$, $\gen{K}$) receive corrections for the amplitude as do for the spin chain. Moreover, the deformed constraint for the amplitude
\[
\bar{\gen{S}}_0 \mathcal{A}^{\text{N}^k\text{MHV}}_n + \bar{\gen{S}}_+  \mathcal{A}^{\text{N}^k\text{MHV}}_{n-1}=0
\]
closely resembles the constraint on the dilatation generator
\[
\comm{\dot{\gen{S}}_{1 \rightarrow 1}}{\gen{D}_{L \rightarrow 2}} + \comm{\dot{\gen{S}}_{2 \rightarrow 1}}{\gen{D}_{L-1 \rightarrow 2}}=0. \label{eq:examplesdotcommuatator}
\]
With this motivation, we take a preliminary straightforward step of translating the spin chain letters into the  scattering amplitude spinor helicity superspace coordinates, which we abbreviate using $\Lambda$. This translation has been considered previously in \cite{Beisert:2010jq}. In the spinor helicity language, states become polynomials in $\Lambda$,  and the dilatation generator gives a map between polynomials.   We then propose that the $L \rightarrow 2$ dilatation generator interactions equal the amplitude operator $\mathbf{A}$: 
\[ 
\langle \Lambda_1 \Lambda_2 | \gen{D}_{L \rightarrow 2} | \mathcal{P}_L \rangle = \langle \Lambda_1 \Lambda_2 |\mathbf{A}|\mathcal{P}_L\rangle 
= c_L \int \mathrm{d} \Lambda' A_{L+2}(\Lambda_1^-,\Lambda_2^-,\Lambda'_1, \ldots \Lambda_L') \mathcal{P}_L(\Lambda'). \label{eq:dequalsa}
\]
Here $\mathcal{P}_L$ is a spinor helicity superspace polynomial representing a $L$-site state. $\gen{D}_{L \rightarrow 2}$ maps this to a new polynomial in two variables, which is then evaluated at $(\Lambda_1, \Lambda_2)$.   $\mathbf{A}$ has matrix elements equal to the sum over $k$ of all $\text{N}^k\text{MHV}$ $(L+2)$-particle tree amplitudes, with the first two particles having negative energy. As we will show, this relation between scattering amplitudes and the dilatation generator (\ref{eq:dequalsa}) ensures that superconformal constraints such as (\ref{eq:examplesdotcommuatator}) are satisfied. Explicit evaluation for $L=2,3$ provides confirmation of (\ref{eq:dequalsa}), though for $L=2$ a simple regularization is required.  An exciting possibility is that (leading) dilatation generator interactions $\gen{D}_{L \rightarrow L'}$ with $L,L' > 2$  also can be  related to scattering amplitudes, though perhaps in a more subtle way. Aside from some comments in the concluding section, we leave this question for future investigation.

We begin in Section \ref{sec:themodel} by discussing basic properties of the spin chain model and reviewing the one-loop dilatation generator and the leading corrections to $\gen{S}$, $\dot{\gen{S}}$, and $\gen{K}$. A construction of $\gen{D}_{L \rightarrow 2}$ is given in Section \ref{sec:theconstruction}. Section  \ref{sec:translation} gives the translation to spinor helicity superspace, and in Section \ref{sec:theproposal} we describe the main relation (\ref{eq:dequalsa}) in more detail and check it for $L=2$ and for $L=3$. We prove superconformal symmetry in Section \ref{sec:superconformal}, but as discussed there, we do rely on numerical confirmation for one special small $L$ case, and our proof assumes that (\ref{eq:dequalsa}) maps polynomials to polynomials for all $L$ without divergences (other than the one that is easily regularized for $L=2$). We discuss further directions for research  in the final section.

\section{The spin chain model \label{sec:themodel}}

This section reviews the formulation of the $\mathcal{N}=4$ SYM spectral problem in terms of spin chain language and discusses structural properties of the perturbative corrections to the spin chain representation. For a more complete description of the basic model see \cite{Beisert:2004ry}.  Section \ref{sec:leadingorder} reviews the classical linear $\alg{psu}(2,2|4)$ representation, while Section \ref{sec: beyondleadingorder} discusses  multisite corrections to the superconformal symmetry generators. Section \ref{sec:harmonic} reviews the harmonic action of the one-loop dilatation generator  \cite{Beisert:2003jj} and a generalization of the harmonic action that gives simple expressions for leading corrections to the supersymmetry generators\cite{Zwiebel:2007th}. Finally, Section \ref{sec:multisitecommutators} examines the structural properties of commutators of multisite interactions and reviews generalized gauge transformations. Appendix \ref{section:appendixgaugetransformations}  gives new explicit results for the gauge transformations that are relevant for this work.

\subsection{The classical linear $\alg{psu}(2,2|4)$ representation \label{sec:leadingorder}}

Single-trace local operators of $\mathcal{N}=4$ SYM correspond to tensor products of  field-strength multiplet elements.  This multiplet includes covariant derivatives acting on the scalars, fermions, or field strength components of $\mathcal{N}=4$ SYM. We will use an oscillator representation for the field strength multiplet \cite{Gunaydin:1984fk}, which uses two doublets of bosonic oscillators $\mathbf{a}^\alpha$ and $\mathbf{b}^{\dot{\alpha}}$  and four fermionic  $\mathbf{d}^A$. The nonvanishing commutation relations are
\[
\comm{\mathbf{a}_\alpha}{\mathbf{a}^{\dagger \beta}} = \delta_\alpha^\beta, \quad \comm{\mathbf{b}_{\dot{\alpha}}}{\mathbf{b}^{\dagger\dot{\beta}}} = \delta_{\dot{\alpha}}^{\dot{\beta}}, \quad \acomm{\mathbf{d}_A}{\mathbf{d}^{\dagger B}} = \delta_A^B.
\]
Then a general multiplet element is
\[ \label{eq:introducevectornotation}
\state{\vec{n}} = \prod_{\alpha=1, 2}(\mathbf{a}^{\dagger \alpha})^{n_\alpha} \prod_{\dot{\beta}=1, 2}(\mathbf{b}^{\dagger \dot{\beta}})^{n_{2+\dot{\beta}}} \prod_{C=1}^4(\mathbf{d}^{\dagger C})^{n_{4+C}}\state{0}.
\]
We label multiplet elements using a vector $\vec{n}=(n_1,n_2,\ldots n_8)$, where the $\mathbf{a}^\dagger$ and $\mathbf{b}^\dagger$ excitation numbers  $(n_1,n_2)$ and $(n_3,n_4)$ are nonnegative integers, and the $\mathbf{d}^\dagger$ excitation numbers $(n_5,n_6, n_7, n_8)$ are 0 or 1. As explained below, additionally the $n_i$ must satisfy
\[ \label{eq:centralchargecondition}
n_1+n_2+2 = n_3+n_4+n_5+n_6+n_7+n_8.
\]
We then consider tensor products of the individual $\state{\vec{n}}$ that are identified under cyclic shifts,  
\[
\state{\vec{n}_1, \vec{n}_2 \ldots \vec{n}_L},
\]
This gives a complete basis for general states (single-trace local operators of planar $\mathcal{N}=4$). We define an inner product by\footnote{Note that this inner product does not include a summation over cyclic permutations. Later, we will extend the superconformal algebra to include generalized gauge transformations, and this modified algebra will be satisfied ``locally''  (without taking into account cyclic identification). Since generalized gauge transformations vanish on cyclic states, the restriction to cyclic states then gives a representation of the undeformed superconformal algebra. This inner product defined without summation over cyclic permutations is the more useful inner product for this approach.}
\[ \label{eq:innerproduct}
\langle \vec{n}_1 \ldots \vec{n}_{L'} | \vec{m}_1 \ldots \vec{m}_L \rangle = \delta_{L L'} \prod_{i=1}^L \delta(\vec{n}_i-\vec{m}_i),
\]
where $\delta((0,0,0,0,0,0,0,0))$ is one, and otherwise $\delta(\vec{n})$ is zero. When oscillators act on a multisite state, we will include a site index subscript $i$ as $\mathbf{a}^{\dagger\alpha}_i$ or $\mathbf{a}_{i,\alpha}$.

The superconformal symmetry of $\mathcal{N}=4$ SYM implies that the states defined above transform under $\alg{psu}(2,2|4)$. In other words, the symmetry generators $\gen{J}$ of $\alg{psu}(2,2|4)$  act on the states and satisfy 
\[
\gcomm{\gen{J}^A}{\gen{J}^B} = f^{AB}{}_C \, \gen{J}^C,
\]
where $ f^{AB}{}_C $ are the $\alg{psu}(2,2|4)$ structure constants.  Two additional $\alg{u}(1)$ generators, $\gen{B}$ and the central charge $\gen{C}$ will also be important. At the classical linear level, the action of the $\gen{J}$  and  $\gen{B}$ and $\gen{C}$ can be written simply in terms of oscillators, 
\[ \label{eq:definepsu224oscillator}
\begin{array}{r@{=}lr@{=}lr@{=}l}
\gen{L}^\alpha_\beta & \mathbf{a}^{\dagger \alpha} \mathbf{a}_\beta -\half \delta^\alpha_\beta n_{\mathbf{a}},  & \gen{Q}^{\alpha B}&\mathbf{a}^{\dagger \alpha} \mathbf{d}^{\dagger B},& \gen{P}^{\alpha \dot{\beta}} & \mathbf{a}^{\dagger \alpha} \mathbf{b}^{\dagger \dot{\beta}}, \\
\dot{\gen{L}}^{\dot{\alpha}}_{\dot{\beta}} & \mathbf{b}^{\dagger \dot{\alpha}} \mathbf{b}_{\dot{\beta}} -\half \delta^{\dot{\alpha}}_{\dot{\beta}}  n_{\mathbf{b}},  &  \gen{S}_{\alpha B}&\mathbf{a}_\alpha \mathbf{d}_B, & \gen{K}_{\alpha \dot{\beta}} & \mathbf{a}_\alpha \mathbf{b}_{\dot{\beta}}, \\
\gen{R}^A_B &\mathbf{d}^{\dagger A}\mathbf{d}_B -\sfrac{1}{4} \delta^A_B n_{\mathbf{d}}, &  \dot{\gen{Q}}^{\dot{\alpha}}_B&\mathbf{b}^{\dagger \dot{\alpha}} \mathbf{d}_ B, & \gen{C}&\half  (n_{\mathbf{a}}-n_{\mathbf{b}}-n_{\mathbf{d}}) + 1, \\
\gen{D}& \half (n_{\mathbf{a}} + n_{\mathbf{b}})  + 1,  & \dot{\gen{S}}_{\dot{\alpha}}^B&\mathbf{b}_{\dot{\alpha}} \mathbf{d}^{\dagger B},  & \gen{B} &n_{\mathbf{d}}.
\end{array}
\]
$n_{\mathbf{a}} =  \mathbf{a}^{\dagger \gamma} \mathbf{a}_\gamma $ counts  $\mathbf{a}$ oscillators, and similarly for $n_{\mathbf{b}}$ and $n_{\mathbf{d}}$.
As usual, the symmetry generators' action on states is the sum of their action on individual sites. Now we recognize the condition on the $n_i$ (\ref{eq:centralchargecondition}) as the statement that the field-strength multiplet elements have zero central charge ($\gen{C}=0$). It is straightforward to translate (\ref{eq:definepsu224oscillator}) to the vector notation of (\ref{eq:introducevectornotation}).
For instance, we have
\[
\gen{L}^\alpha_\beta \state{\vec{n}} = (n_{\beta}- \half \delta^\alpha_\beta (n_1+n_2)) \state{\vec{n} + \vec{\delta}_\alpha- \vec{\delta}_\beta}.
\]
$\vec{\delta}_{\alpha}$ is an eight-dimensional vector with zeros in all entries beside the $\alpha$-th entry, which is one. Similarly, $\vec{\delta}_{\alpha \beta \ldots}$  has one in entries $\alpha, \beta, \ldots$, and remaining entries zero.

We also define $\mathbf{c}^A=\mathbf{d}^{\dagger A}$, $\mathbf{c}_A^\dagger=\mathbf{d}_A$, so $n_{\mathbf{c}} = (4 - n_{\mathbf{d}})$ (for a single site).
Then the representation transforms simply under the conjugation transformation\footnote{
In Minkowski space, $\mathbf{a}$ and $\mathbf{b}$ would need to be complex conjugates of each other. While we will not impose this constraint since it is not necessary for the algebraic approach of this work, the symmetry under this complex conjugation transformation remains.}
\[ \label{eq:conjugation}
 \mathbf{a} \leftrightarrow \mathbf{b} , \quad \mathbf{d} \leftrightarrow \mathbf{c}.
 \]
For example, this interchanges $\gen{L}$ and $\dot{\gen{L}}$, $\gen{Q}$ and $\dot{\gen{Q}}$ and $\gen{S}$ and $\dot{\gen{S}}$, but $\gen{D}$ is invariant. This  conjugation (\ref{eq:conjugation}) changes states as 
 \[
 \vec{n} \leftrightarrow \vec{n}^\ast, \quad \vec{n}^\ast=(n_3,n_4,n_1,n_2,(1-n_5),(1-n_6),(1-n_7),(1-n_8)).
 \]
Note that $\vec{n}^\ast$ has zero central charge if $\vec{n}$ does, and the $\gen{B}$-charges $B$ and $B^\ast$ satisfy $B^\ast = 4 - B$. 

Also, we can define Hermitian conjugation for the superconformal generators by considering radial quantization of $\mathcal{N}=4$ SYM (where the dimensions of operators become energies of state on $S^3$). Here Hermitian conjugation simply interchanges creation and annihilation oscillators. For example, $\gen{Q}^\dagger$= $\gen{S}$, and $\gen{D}^\dagger = \gen{D}$.

\subsection{Beyond the linear representation \label{sec: beyondleadingorder}}

The $\alg{psu}(2,2,|4)$ representation described in the previous section is only valid when $g=0$\footnote{This assumes a standard coupling constant convention. As we will see, with the coupling constant convention used in this work the classical linear representation is deformed even at $g=0$.}. When we expand perturbatively in $g$, the representation of superconformal symmetry becomes deformed. However, the  $\alg{psu}(2,2|4)$ algebra is undeformed:
\[
\gcomm{\gen{J}^A(g)}{\gen{J}^B(g)} = f^{AB}_C \gen{J}^C (g).
\]
The deformed symmetry generators $\gen{J}^A(g)$ include interactions that act on multiple sites at once, and interactions that change the number of sites. Since deformations of the symmetry generators are central to this work, here we includes a discussion of properties of these multisite deformation, including their coupling constant expansion in our normalization, their classification according to oscillator number and length changes, and their discrete symmetries.

We use the notation $\gen{J}_{L \rightarrow L'}$ to describe interactions that act simultaneously on $L$ adjacent\footnote{For finite-rank gauge group, $\gen{J}$ is a sum of connected interactions. Here connectedness refers to the contractions of the gauge group indices. In the planar limit this implies that $\gen{J}$ acts on adjacent sites.} sites and change the length by $(L'-L)$. In other words,  $\gen{J}_{L \rightarrow L'}$  is a linear combination of basic interactions that remove $L$ adjacent $\vec{m}$ and replace them with $L'$ adjacent $\vec{n}$. Of course, $\gen{J}_{L \rightarrow L'}$ only acts (nontrivially) on spin chains with at least $L$ sites.

Next we find the coupling constant dependence for the multisite interactions. A Feynman diagram expansion of $\gen{J}_{L \rightarrow L'}$ implies  \cite{Beisert:2004ry}
\[ \label{eq:loopexpansion}
\gen{J}_{L \rightarrow L'}(g) = \sum_{l=0}^\infty g^{L+L'+2 l - 2} \gen{J}^{(l)} _{L \rightarrow L'},
\]
 where $l$ corresponds to the number of Feynman diagram momentum loops\footnote{As (\ref{eq:loopexpansion}) shows, and as explained in  \cite{Beisert:2004ry}, powers of $g$ for a Feynman graph contribution to $\gen{J}_{L \rightarrow L'}$ arise both from momentum loops and from connecting components of the graph. Note that by definition, the ``m-loop dilatation generator'' is the $\mathcal{O}(g^{2 m})$ part, so that this ``loop'' order  counts both momentum loops and components connected in Feynman graphs.}. In this work we will restrict to $l=0$. For the remainder of this work we use $\gen{J}_{L \rightarrow L'}$ to mean the $l=0$ term $\gen{J}^{(0)}_{L \rightarrow L'}$. It is convenient to perform a similarity transformation with respect to the length operator $\mathbf{L}$, which counts the number of spin-chain sites, as 
 \[
 \gen{J} \rightarrow g^{\mathbf{L}} \, \gen{J} \, g^{-\mathbf{L}} \quad  \Rightarrow \quad \gen{J}_{L \rightarrow L'} \rightarrow g^{L'} \, \gen{J} \, g^{-L}.
 \]
 Importantly, because this is a similarity transformation it maps one representation of superconformal symmetry to an equivalent one that has the same eigenvalues for the dilatation generator.  Applying the restriction to $l=0$ and the similarity transformation, the coupling constant dependence of (\ref{eq:loopexpansion}) simplifies to 
\[
\gen{J}_{L \rightarrow L'}(g) =  g^{2 L' - 2} \gen{J} _{L \rightarrow L'}, \label{eq:simplercouplingdependence}
\]
which expands in non-negative even integer powers of $g$. Due to this similarity transformation, $\gen{D}_{L \rightarrow 2}$ now appears at $\mathcal{O}(g^2)$, but $\gen{D}_{2 \rightarrow L}$ appears at $\mathcal{O}(g^{2 L -2})$. In the conventional normalization, these would instead both appear at ``$(L/2)$-loops'' ($\mathcal{O}(g^{L})$).
 
The most important deformation is the anomalous part of the dilatation generator, $\delta \gen{D}(g)= \gen{D}(g)-\gen{D}_0$. The eigenvalues of $\delta \gen{D}$ give the spectrum of anomalous dimensions of $\mathcal{N}=4$ SYM. As usual, we use a renormalization scheme satisfying
\[
\comm{\gen{D}_0}{\gen{J}(g)}= \mathrm{dim}(J) \, \gen{J}(g), \label{eq:dimpreserved}
\]
where $\mathrm{dim}(J)$ is the classical engineering dimension of $\gen{J}$ and can be inferred from the classical linear representation of (\ref{eq:definepsu224oscillator}).  This yields the very useful simplification that $\delta \gen{D}(g)$ is a $\mathbb{R}$ generator.
 
Next we will classify the possible oscillator number of changes for the multisite interactions. This classification gives significant constraints on the corrections to the superconformal generators. First note that, since Lorentz and $R$-symmetry are manifest, $\gen{L}$, $\dot{\gen{L}}$ and $\gen{R}$ receive no corrections. However, all of the other $\alg{psu}(2,2|4)$ are corrected by multisite interactions. 
Combining (\ref{eq:dimpreserved}) with Lorentz and $R$-symmetry and the central charge constraint (\ref{eq:centralchargecondition}), it is straightforward to find basic building blocks for the oscillator number changes of length-changing interactions. For this we define two new operators, $\mathbf{L}^\pm_i$ that increase or decrease the length of spin chain states by one site. $\mathbf{L}^+_i$ inserts site $i$ with no oscillator excitations, and leaves the oscillator exciations on the original sites unchanged. Its Hermitian conjugate $\mathbf{L}^-_i$ annihilates a state unless site $i$ has no oscillator excitations, in which case it returns the state with site $i$ removed and with no changes to the oscillators of the other sites.  There are two length-decreasing building blocks that remove one site and change oscillator numbers as
\[
\varepsilon_{\alpha \beta} \mathbf{a}_i^{\dagger \alpha} \mathbf{a}_j^{\dagger \beta} \mathbf{L}^-, \quad \quad
 (\varepsilon_{\dot{\alpha} \dot{\beta}} \mathbf{b}_i^{\dagger\dot{\alpha}}  \mathbf{b}_j^{\dagger\dot{\beta}}) \,( \varepsilon^{A B C D} \, \mathbf{d}_A \mathbf{d}_B \mathbf{d}_C \mathbf{d}_D)\mathbf{L}^- . \label{eq:basicbuildingblocks}
\]
Here we are being schematic and just keeping track of total oscillator number changes. Therefore,  we suppressed the site index for $\mathbf{L}^-$ and $\mathbf{d}$. However, we still included the site indices $i,j$ on $\mathbf{a}^\dagger$ and  $\mathbf{b}^\dagger$ to emphasize that these bosonic oscillators cannot be antisymmetrized on a single site, which will be an important constraint. In contrast, the four $\mathbf{d}$ of the second interaction could act on the same site (or on different sites). There are also two basic building blocks  that increase the number of sites by one that are given by the Hermitian conjugates of (\ref{eq:basicbuildingblocks}),
\[
\varepsilon^{\alpha \beta} \mathbf{a}_{i,\alpha} \mathbf{a}_{j,\beta}\mathbf{L}^+, \quad \quad
 (\varepsilon^{\dot{\alpha} \dot{\beta}} \mathbf{b}_{i,\dot{\alpha}} \mathbf{b}_{j,\dot{\beta}}) \, ( \varepsilon_{A B C D} \, \mathbf{d}^{\dagger A} \mathbf{d}^{\dagger A} \mathbf{d}^{\dagger C} \mathbf{d}^{\dagger D})\mathbf{L}^+ . \label{eq:basicbuildingblocksconjugate}
\]
 The general $\gen{J}_{L \rightarrow L'}$ interaction then includes $p\geq 0$ length-decreasing building blocks (\ref{eq:basicbuildingblocks}) and $(L'-L+p)\geq 0$  length-increasing building blocks (\ref{eq:basicbuildingblocksconjugate})\footnote{To avoid redundancy, only one type of each pair of Hermitian conjugate building blocks should appear, since Hermitian conjugate building blocks cancel each other.}, as well as the oscillator number changes of $\gen{J}_0$. These oscillator creations/annihilations are then combined with a permutation of oscillators across the final $L'$ sites.

In addition to the $l=0$ restriction above, in the remainder of this work we will also restrict to the leading order in $g$, using the convention of (\ref{eq:simplercouplingdependence}). According to (\ref{eq:simplercouplingdependence}) the leading $\alg{psu}(2,2|4)$ generator interactions are $\mathcal{O}(g^0)$ and have a single final site. However, the anomalous part of the dilatation generator, $\delta \gen{D}$ begins at order $g^2$, and so at leading order we also include all interactions of $\delta \gen{D}$ with two final sites. From our above analysis of the basic building blocks of length-changing interactions, we immediately conclude that $\gen{Q}$, $\dot{\gen{Q}}$ and $\gen{P}$ have no corrections at $\mathcal{O}(g^0)$ (because a $\alg{su}(2)$ singlet of bosonic oscillators requires more than one final site). On the other hand, at $\mathcal{O}(g^0)$ the interactions
\[
(\gen{S}_{\alpha B})_{2 \rightarrow 1} \sim \varepsilon_{\alpha \beta} \mathbf{a}^{\dagger \beta} \mathbf{d}_B, \quad (\dot{\gen{S}}_{\dot{\alpha}}^B)_{2 \rightarrow 1} \sim \varepsilon_{\dot{\alpha}\dot{ \beta}} \mathbf{b}^{\dagger \dot{\beta}} \, \varepsilon^{BCDE} \mathbf{d}_C\mathbf{d}_D\mathbf{d}_E
\]
are possible. However, $\gen{S}$ and $\dot{\gen{S}}$ have  no $L \rightarrow 1$ interactions for $L > 2$.  Since $\gen{S}$ and $\dot{\gen{S}}$ anticommute to $\gen{K}$, it follows that $\gen{K}$ has  $2 \rightarrow 1$ and $3 \rightarrow 1$ corrections, but no other corrections at this order. For $\delta \gen{D}$, $L \rightarrow 2$ interactions occur for all $L \geq 2$, but $L \rightarrow 1$ interactions for $L \geq 2$ are not possible, since these would required antisymmetrizing $\mathbf{a}^\dagger$ or $\mathbf{b}^\dagger$ on  a single site\footnote{The $1 \rightarrow 1$ interactions of the dilatation generator are its classical part, as well as loop corrections that we do not study in this work.}. In summary, for the restriction to leading order of this work we have
\<
\gen{S}_{\alpha B} \eq (\gen{S}_{\alpha B})_{1 \rightarrow 1} + (\gen{S}_{\alpha B})_{2 \rightarrow 1}, \nln
\dot{\gen{S}}_{\dot{\alpha}}^B \eq (\dot{\gen{S}}_{\dot{\alpha}}^B)_{1 \rightarrow 1} + (\dot{\gen{S}}_{\dot{\alpha}}^B)_{2 \rightarrow 1}  ,
\nln
\gen{K}_{\alpha \dot{\beta}} \eq (\gen{K}_{\alpha \dot{\beta}})_{1 \rightarrow 1} + (\gen{K}_{\alpha \dot{\beta}})_{2 \rightarrow 1} + (\gen{K}_{\alpha \dot{\beta}})_{3 \rightarrow 1} ,
\nln
\delta \gen{D} \eq g^2 \sum_{L=2}^\infty \gen{D}_{L \rightarrow 2}.
\>
At this order, all other $\alg{psu}(2,2|4)$ generators have only $1 \rightarrow 1$ interactions, as do  $\gen{B}$ and $\gen{C}$.

We label interactions by $B$-charge as
\[ \label{eq:bchargelabel}
\comm{\gen{B}}{\gen{J}^{[k]}_{L \rightarrow L'}} = -4(L-L'-k)\gen{J}^{[k]}_{L \rightarrow L'},
\]
which means that $\gen{J}^{[k]}_{L \rightarrow L'}$ removes $4(L-L'-k)$ $\mathbf{d}$ oscillators. The $2 \rightarrow 1$ interactions for the supercharges are then  $\gen{S}^{[1]}_{2 \rightarrow 1}$ and $\dot{\gen{S}}^{[0]}_{2 \rightarrow 1}$. It follows that $\gen{K}_{2 \rightarrow 1}$ has  components $\gen{K}^{[0]}_{2 \rightarrow 1}$ and $\gen{K}^{[1]}_{2 \rightarrow 1}$, while the $3 \rightarrow 1$ interaction has only one  component $\gen{K}^{[1]}_{3 \rightarrow 1}$. The $B$-charge expansion for $\delta \gen{D}$ is
\[ \label{eq:bchargeexpansion}
\delta \gen{D} = g^2 \sum_{L=2}^{\infty} \sum_{k=0}^{L-2} \gen{D}^{[k]}_{L \rightarrow 2}.
\]

  Like the  $\gen{J}_{1 \rightarrow 1}$ of the $g=0$ representation, the $\gen{J}_{L \rightarrow L'}$ actions  are summed over the length of the spin chain as 
\[ \label{eq:sumoverspinchain}
\gen{J}_{L \rightarrow L'}\state{\vec{m}_1 \ldots \vec{m}_M} = \sum_{i=1}^M \gen{J}_{L \rightarrow L'}(i,i+1, \ldots i+L-1) \state{\vec{m}_1 \ldots \vec{m}_M}.
\]
The argument  of $\gen{J}_{L \rightarrow L'}$,  specifies the initial sites acted on, and we use periodic identification of spin chain sites\footnote{Using a shift operator simplifies the sum over spin chain sites, but we will not need that in this work.}.   
Most of the time it will be sufficient to just think about the interaction density, that is the interaction acting on a single set of $L$ adjacent sites, which we can specify simply by giving the action of $\gen{J}_{L \rightarrow L'}(1, \ldots L)$. 
We will also frequently look at matrix elements instead. Because the generators act homogeneously along the spin chain,  the set of matrix elements, for example,
 \[
 \langle \vec{n}_1\vec{n}_2 | \gen{J}_{L \rightarrow 2} \state{\vec{m}_1, \vec{m}_2 \ldots \vec{m}_L}
 \]
 completely specify $\gen{J}_{L \rightarrow 2}$. Here we use the inner product defined in (\ref{eq:innerproduct}). 
 
 We conclude this sections by introducing three discrete symmetries that we will use to shorten the construction of corrections to the dilatation generator. Due to charge conjugation symmetry of $\mathcal{N}=4$ SYM, all $\alg{psu}(2,2|4)$ generators are spin chain parity even. We will use parity to compute $L \rightarrow 2$ interactions, where it gives the condition
 \[
 \langle\vec{n}_1\vec{n}_2|\gen{J}_{L \rightarrow 2}|\vec{m}_1  \ldots \vec{m}_L\rangle = (-1)^{L + (n_{1,f})(n_{2,f}) + \prod_{i<j}( m_{i,f})(m_{j,f})}  \langle\vec{n}_2\vec{n}_1|\gen{J}_{L \rightarrow 2}|\vec{m}_L \vec{m}_{L-1}  \ldots \vec{m}_1\rangle. \label{eq:parity} \]
 $n_{i,f}$ is the number of fermionic oscillators on the $i$th site.
 
 The dilatation generator has two other discrete symmetries. First, $\gen{D}$ is symmetric under conjugation (\ref{eq:conjugation}), 
\[
\langle\vec{n}_1\vec{n}_2|\gen{D}^{[k]}_{L \rightarrow 2}|\vec{m}_1  \ldots \vec{m}_L\rangle = \langle\vec{n}^\ast_1\vec{n}^\ast_2|\gen{D}^{[L-2-k]}_{L \rightarrow 2}|\vec{m}^\ast_1  \ldots \vec{m}^\ast_L\rangle.
\]
Second, due to Hermiticity there is a basis in which the matrix elements of $\gen{D}$ and $\gen{D}^\dagger$ are equal\footnote{This is expected to hold at all orders in perturbation theory with respect to the undeformed scalar product (\ref{eq:innerproduct}), but we are not aware of a rigorous proof that the relevant scalar product remains undeformed.  (\ref{eq:hermitianconjugationofd}) holds (with respect to the undeformed scalar product) at least at leading order because of the Hermiticity of the (leading) superconformal generators that algebraically  fix  the leading $\gen{D}_{L \rightarrow 2}$ and  $\gen{D}_{2 \rightarrow L}$.}. For our basis,
 \< \label{eq:hermitianconjugationofd}
\langle\vec{m}_1  \ldots \vec{m}_L|\gen{D}_{2 \rightarrow L}|\vec{n}_1\vec{n}_2 \rangle \eq \frac{\vec{n}_1! \,\vec{n}_2!}{\vec{m}_1! \,\vec{m}_2! \ldots \vec{m}_L!} \langle\vec{n}_1\vec{n}_2|\gen{D}_{L \rightarrow 2}|\vec{m}_1  \ldots \vec{m}_L\rangle  , \nln
\vec{n}_i! \eq n_{i,1}! \, n_{i,2}! \, n_{i,3}! \, n_{i,4}!
\>
Note that with our coupling constant convention the leading interactions for $\gen{D}_{2 \rightarrow L}$ have coefficient $g^{2 L -2}$.

\subsection{The one-loop dilatation generator and the corrections to $\gen{S}$, $\dot{\gen{S}}$, $\gen{K}$  \label{sec:harmonic}}

Our construction for $\gen{D}_{L \rightarrow 2}$ will start from the known one-loop dilatation generator, so we present the one-loop dilatation generator in this section. Here we also give the corrections to the other superconformal generators, which will be important later both for the construction of $\gen{D}_{L \rightarrow 2}$ and for proving that the proposed relation to scattering amplitudes is consistent with superconformal symmetry. 

 First we give an expression for $\gen{D}_{2 \rightarrow 2}$, the one-loop dilatation generator of $\mathcal{N}=4$ SYM, in terms of the ``harmonic action'' \cite{Beisert:2003jj}.  We use $\mathbf{A}^{a \dagger}$ with superscript index $a=1,2\ldots 8$  to represent the eight flavors of oscillators, $(\mathbf{a}^{\alpha \dagger},\mathbf{b}^{\dot{\beta} \dagger}, \mathbf{d}^{C \dagger})$. Then  a general two-site state is written as
 \[
 \state{p_1,p_2 \ldots p_n;A} = \mathbf{A}^{a_1 \dagger}_{p_1} \mathbf{A}^{a_2 \dagger}_{p_2} \ldots \mathbf{A}^{a_n \dagger}_{p_n}\state{\vec{d} \, \vec{d}},
 \]
 where $p_i$ is the site index 1 or 2 and $\state{\vec{d}\, \vec{d}}$ has no excited oscillators (the vacuum with respect to $\mathbf{d}$ oscillators).  Introducing a slight generalization of the original harmonic action, we define a harmonic action generator $\mathcal{H}^{(r,s)}$ which gives a weighted sum over shifts of oscillators between sites as 
 \[ \label{eq:defineharmonicaction}
 \mathcal{H}^{(r,s)}   \state{p_1,p_2 \ldots p_n;A} =\sum_{p_1',p_2'\ldots p_n'} c^{(r,s)}(n, n_{12}, n_{21})  \state{p_1',p_2' \ldots p_n';A}.
 \]
 Each $p_i'$ is summed from 1 to 2. For each term in the sum, $n_{12}$ is the number of oscillators that shift from site 1 to 2, and $n_{21}$ is the number of oscillators that shift from site 2 to 1. Note that $n$ is the total number of oscillators of the initial state, and the total number of oscillators is unchanged by $ \mathcal{H}^{(r,s)}$. The  weight function is given by\footnote{The sign factor $ (-1)^{n_{21}}$ could also be written in terms of $n_{12}$. Since physical states with zero central charge have an even number of oscillators,  for $\gen{D}_{2 \rightarrow 2}$ $n_{12}+n_{21}$ must be even, and we can use  $ (-1)^{n_{12}}$. In the expressions (\ref{eq:defineinteractingsupercharge}) for $\gen{S}$ and $\dot{\gen{S}}$,   $\mathcal{H}^{(r,s)}$ maps two states with zero central charge to two states with half-integer central charge, so there we can use $ (-1)^{1+n_{12}}$ instead.}
\[ 
 c^{(r,s)}(n, n_{12}, n_{21}) = (-1)^{n_{21}} B(\half (n_{12}+n_{21})+r, \half(n-n_{12}-n_{21}) + s).
\]
 Here $B$ is the  Euler beta function $B(a,b) = \Gamma(a)\Gamma(b)/\Gamma(a+b)$.  We then have 
\[
\gen{D}_{2 \rightarrow 2} = -2\, \delta_{C_1,0}\mathcal{H}^{(0,1)}_{\text{regularized}}, \label{eq:D2to2}
\]
where $\delta_{C_1,0}$ enforces the zero central charge condition on the first site\footnote{This is sufficient because the harmonic action preserves the total central charge.}. We regularize the coefficient for  $n_{12}=n_{21}=0$ as  $c^{(0,1)}(n, 0, 0)= -S_1(n)$. $S_1(n)$ is the $n$th ordinary harmonic number, 
\[
S_1(n) = \sum_{k=1}^n \frac{1}{k}.
\]
For later use we introduce $S_m(n)$, the $n$th Harmonic number of order $m$,  which is given by
\[ \label{eq:defineharmonics}
S_m(n) = \sum_{k=1}^n \frac{1}{k^m}.
\]

$\mathcal{H}^{(r,s)}$ also makes possible compact expressions for $\dot{\gen{S}}_{2 \rightarrow 1}$ and $\gen{S}_{2 \rightarrow 1}$. In terms of  matrix elements  \cite{Zwiebel:2007th},
\<
\langle\vec{n} | (\dot{\gen{S}}_{\dot{\alpha}}^C)_{2 \rightarrow 1}\state{\vec{m}_1\vec{m}_2} \eq  \langle\vec{n} \, \vec{c}| \varepsilon_{\dot{\beta}\dot{\alpha}}\mathbf{b}_1^{\dagger\dot{\beta}} \mathbf{d}_{2}^{\dagger C} \mathcal{H}^{(\half,\half)}\state{\vec{m}_1\vec{m}_2},
\nln
\langle\vec{n} | (\gen{S}_{\alpha C})_{2 \rightarrow 1}\state{\vec{m}_1\vec{m}_2} \eq  \langle\vec{n}\, \vec{d}| \varepsilon_{ \beta \alpha}\mathbf{a}_1^{\dagger \beta} \mathbf{d}_{2,C} \mathcal{H}^{(\half,\half)}\state{\vec{m}_1\vec{m}_2},
\notag \\
\vec{c} = (0,0,0,0,1,1,1,1), &\quad  &  \vec{d} = (0,0,0,0,0,0,0,0).
 \label{eq:defineinteractingsupercharge}
\>
These expressions do not need regularization.

Now the $\gen{K}$ corrections are fixed by the $\alg{psu}(2,2|4)$ algebra as\footnote{The use of the $R$-index of $1$ is just an arbitrary choice.}
\<
(\gen{K}_{\alpha \dot{\beta}})_{2 \rightarrow 1}\eq \acomm{(\gen{S}_{\alpha 1})_{2 \rightarrow 1}}{(\dot{\gen{S}}_{\dot{\beta}}^1)_{1 \rightarrow 1}}+ \acomm{(\gen{S}_{\alpha 1})_{1 \rightarrow 1}}{(\dot{\gen{S}}_{\dot{\beta}}^1)_{2 \rightarrow 1}}, 
\nln
 (\gen{K}_{\alpha \dot{\beta}})_{3 \rightarrow 1} \eq \acomm{(\gen{S}_{\alpha 1})_{2 \rightarrow 1}}{(\dot{\gen{S}}_{\dot{\beta}}^1)_{2 \rightarrow 1}}.
\>
It is straightforward to prove that these corrections to the supercharges and $\gen{K}$ are consistent with superconformal symmetry. In fact, like the one-loop dilatation generator, they are fixed (up to normalization) by superconformal symmetry \cite{Zwiebel:2007th}.
 
Finally, the generalized harmonic action can be written in a very simple form \cite{Zwiebel:2007th}. We include this form especially because it is well-suited for making the connection to scattering amplitudes. Again using the notation $\mathbf{A}^{a \dagger}_i$ to represent the oscillator of flavor $a=1,2\ldots8$ acting on site $i$, we have
\<
\mathcal{H}^{(r,s)} \prod_{a,b} \mathbf{A}^{a \dagger}_1 \mathbf{A}^{b\dagger}_2 \state{0,0} \eq 2 \int_0^{\pi/2} \mathrm{d} \theta \, (\sin \theta)^{2 r - 1} ( \cos \theta)^{2 s - 1}  \prod_{a,b} \mathbf{A}^{a\dagger}_{1'} \mathbf{A}^{b\dagger}_{2'} \state{0,0},
\nln
\mathbf{A}^{a\dagger}_{1'} \eq \hspace{.3cm} \mathbf{A}^{a\dagger}_{1} \cos \theta  + \mathbf{A}^{a\dagger}_{2} \sin \theta ,
\nln
\mathbf{A}^{a\dagger}_{2'} \eq -\mathbf{A}^{a\dagger}_{1} \sin \theta + \mathbf{A}^{a\dagger}_{2} \cos \theta . \label{eq:Hisrotatingoscillators}
\>
This works because of the elementary identity
\[
2 \int_0^{\pi/2} \mathrm{d} \theta \, (\cos \theta)^{2 p -1}  (\sin \theta)^{2 q -1} = B(p,q).
\]

For the case of $\gen{D}_{2 \rightarrow 2}$ the regularization can be included through the addition of a diagonal term
\[
\gen{D}_{2 \rightarrow 2} \prod_{a,b} \mathbf{A}^{a\dagger}_{1} \mathbf{A}^{b\dagger}_{2} \state{0,0} = 2 \, \delta_{C_1,0} \int_0^{\pi/2} \mathrm{d} \theta \, 2 \cot \theta  \prod_{a,b} \Big(  (\mathbf{A}^{a\dagger}_{1} \mathbf{A}^{b\dagger}_{2} - \mathbf{A}^{a\dagger}_{1'} \mathbf{A}^{a\dagger}_{2'} )\state{0,0} \Big) \label{eq:regularizedd2}.
\]
Alternatively, we could write the central charge projection using a phase integral with, for example,  $\mathbf{a}$ oscillators carrying a positive phase and $\mathbf{b}$ and $\mathbf{d}$ carrying the negative phase of the same magnitude.

\subsection{Multisite commutators and generalized gauge transformations \label{sec:multisitecommutators}}
Much of this work follows from $\delta \gen{D}$ generating a $\alg{u}(1)$, or equivalently:
\[
\comm{\gen{J}}{\delta \gen{D}}= 0,
\]
where $\gen{J}$ is any $\alg{psu}(2,2|4)$ generator (or $\gen{C}$). Consequently, in this section we explain the important structural properties of these commutators, especially that in some cases the commutation relations are only satisfied up to ``generalized gauge transformations.'' This section also gives explicit expressions for the commutators of multisite interactions, which involve summations over spin chain sites. 

The commutator of two local (on the spin chain) symmetry generators gives another local quantity. 
Then the vanishing commutator between $\delta \gen{D}$ and the generators $\gen{J}$ with only $1 \rightarrow 1$ leading order interactions is equivalent to the local condition,
\[
 \sum_{i=1}^2 \gen{J}(i) \gen{D}_{L \rightarrow 2}(1, 2, \ldots L)-  \sum_{i=1}^L \gen{D}_{L \rightarrow 2}(1, 2, \ldots L) \gen{J}(i) = 0. \quad (L \geq 2)
\]

In contrast, the commutators with $\gen{S}$, $\dot{\gen{S}}$ and $\gen{K}$  vanish only when summed over all sites of the spin chain, taking into account the cyclic identification of spin chain sites.  However, we can still work with local expressions if we include generalized gauge transformations. These are interactions that are locally nonzero, but annihilate cyclic spin chain states. For $L \rightarrow 2$ interactions, such generalized gauge transformations $\gen{g}$ have matrix elements\footnote{Additional signs are required for fermionic gauge transformations.},
\[
\langle\vec{n}_1\vec{n}_2|\gen{g}_{L \rightarrow 2}| \vec{m}_1 \ldots \vec{m}_L\rangle = \delta(\vec{n}_1-\vec{m}_1) G(\vec{m}_2 \ldots \vec{m}_L;\vec{n}_2) - \delta(\vec{n}_2-\vec{m}_L) G(\vec{m}_1 \ldots \vec{m}_{L-1};\vec{n}_1)
\]
The interactions of $\gen{g}$ include one bystander site. Since all $\alg{psu}(2,2|4)$ generators are parity even we will also required gauge transformations to be parity even.
Then we have the following commutators satisfied locally on the spin chain,
\<
\comm{\gen{K}_{\alpha \dot{\beta}}}{\delta \gen{D}} \eq g^2 \, \gen{k}_{\alpha \dot{\beta}},
\nln
\comm{\gen{S}_{\alpha C}}{\delta \gen{D}} \eq g^2 \, \gen{s}_{\alpha C},
\nln
\comm{\dot{\gen{S}}_{\dot{\beta}}^C}{\delta \gen{D}} \eq g^2 \, \dot{\gen{s}}_{\dot{\beta}}^C,
\>
where we use lower case Gothic letters $\gen{k}$, $\gen{s}$ and $\dot{\gen{s}}$ to denote the generalized gauge transformations. Recall that  the only possible corrections to $\gen{S}$ and $\dot{\gen{S}}$ are $2 \rightarrow 1$.  Then, since the interactions of $\gen{s}$ and $\dot{\gen{s}}$ include one bystander site, the  fermionic gauge transformations have only $3 \rightarrow 2$ interactions\footnote{The detailed argument is as follows. $\comm{\gen{S}_{2 \rightarrow 1}}{\gen{D}^{[k]}_{L \rightarrow 2}}$ is a $(L+1) \rightarrow 2$ interaction that inserts $k$ singlet pairs of $\mathbf{a}^\dagger$ and $(L-2-k)$ singlet pairs of $\mathbf{b}^\dagger$. Assuming this commutator equals a gauge transformation, it must be a sum of interactions with either the first or the last site a bystander. It follows that there would be only one final site on which to insert any singlet pairs of oscillators. Recall that it is impossible to antisymmetrize $\mathbf{a}^\dagger$ (or $\mathbf{b}^\dagger$) on a single site since  $\comm{\mathbf{a}^{\alpha \dagger}}{\mathbf{a}^{\beta \dagger}}=0$. Therefore, gauge transformations are only possible when $k=(L-2-k)=0$, that is for $L=2$, corresponding to $3 \rightarrow 2$ gauge transformations. The argument can be repeated straightforwardly for $\dot{\gen{S}}$.}: $ \gen{s}_{\alpha C}=(\gen{s}_{\alpha C})_{3 \rightarrow 2}$ and  $ \dot{\gen{s}}_{\dot{\beta}}^C=( \dot{\gen{s}}_{\dot{\beta}}^C)_{3 \rightarrow 2}$. Since $\gen{S}$ and $\dot{\gen{S}}$ anticommute to $\gen{K}$, by the Jacobi identity we have
\[ \label{eq:kgaugefromcommutator}
\gen{k}_{\alpha \dot{\beta}}= \acomm{\gen{S}_{\alpha 1}}{ \dot{\gen{s}}_{\dot{\beta}}^1} +\acomm{ \gen{s}_{\alpha 1}}{\dot{\gen{S}}_{\dot{\beta}}^1}.
\]
Therefore, $\gen{k}_{\alpha \dot{\beta}} = (\gen{k}_{\alpha \dot{\beta}})_{3 \rightarrow 2} + (\gen{k}_{\alpha \dot{\beta}})_{4 \rightarrow 2}$. Appendix \ref{section:appendixgaugetransformations} gives explicit expressions for these gauge transformations. As explained there, the gauge transformations' matrix elements can be derived using a $\alg{psu}(2,1|3)$ sector of the spin chain model.

Including the gauge transformation, and organizing by the number of sites acted upon, the vanishing commutator between $\delta \gen{D}$  and $\dot{\gen{S}}$ gives

\< \label{eq:sdeltadcommutator}
\comm{\dot{\gen{S}}}{g^{-2}\, \delta\gen{D}}_{L \rightarrow 2} \eq  \delta_{L3} \dot{\gen{s}}_{3 \rightarrow 2} = \comm{\dot{\gen{S}}_{1 \rightarrow 1}}{\gen{D}_{L \rightarrow 2}} + \delta_{L\neq2} \, \comm{\dot{\gen{S}}_{2 \rightarrow 1}}{\gen{D}_{(L-1) \rightarrow 2}}  \\
\eq   \sum_{i=1}^2 \dot{\gen{S}}_{1 \rightarrow 1}(i) \gen{D}_{L \rightarrow 2}(1, 2, \ldots L)-  \sum_{i=1}^L \gen{D}_{L \rightarrow 2}(1, 2, \ldots L) \dot{\gen{S}}_{1 \rightarrow 1}(i) \notag \\
&+ &\delta_{L\neq2}   \bigg( \!\dot{\gen{S}}_{2 \rightarrow 1}(2,3) \gen{D}_{(L-1) \rightarrow 2}(1, \ldots L-1) + \dot{\gen{S}}_{2 \rightarrow 1}(1,2) \gen{D}_{(L-1) \rightarrow 2}(2,  \ldots L)  \notag \\
&  & \quad - \sum_{i=1}^{L-1} \gen{D}_{(L-1) \rightarrow 2}(1, 2, \ldots (L-1)) \dot{\gen{S}}_{2 \rightarrow 1}(i,i+1) \bigg) \notag \\
& + & \half \delta_{L3} \bigg( \dot{\gen{S}}_{2 \rightarrow 1}(1,2) \gen{D}_{2\rightarrow 2}(1, 2) +  \dot{\gen{S}}_{2 \rightarrow 1}(2,3) \gen{D}_{2\rightarrow 2}(2,3)\bigg), \notag
\>
where $\delta_{L\neq2}=(1-\delta_{L2})$. To make the equations slightly less cluttered, we have suppressed the indices of $\dot{\gen{S}}^C_{\dot{\alpha}}$. According to our convention  sites to the left (lower-numbered) of a length-changing interactions keep their same site number, while sites to the right have their number shifted. For instance,  $\gen{D}_{(L-1) \rightarrow 2}(1, 2, \ldots L-1)$  replaces the sites $(1,2, \ldots (L-1))$ with sites $1,2$, while the site originally labeled $L$ is now labeled 3.  In the above expression, we have just worked out the commutator $L \rightarrow 2$ local density, with spin chain arguments $(1,2, \ldots L)$. The full commutator should be summed over the length of the spin chain as in (\ref{eq:sumoverspinchain}). However, because we included $\dot{\gen{s}}$, this expression is exact even locally (without the sum over the spin chain).  The factor of $\half$ is needed to avoid double counting of the interactions with one bystander site\footnote{Terms where $\dot{\gen{S}}_{2 \rightarrow 1}$ acts  after $\gen{D}_{L \rightarrow 2}$ on the same sites vanish for $L > 2$ because a $L \rightarrow 1$ interaction inserting a $\alg{su}(2)$ singlet of  $\mathbf{a}^\dagger$ or $\mathbf{b}^\dagger$ must be trivial (zero). Note that the  term  $\dot{\gen{S}}_{2 \rightarrow 1}(2,3) \gen{D}_{(L-1)\rightarrow 2}(1,\ldots (L-1))$ is not of this form since after $\gen{D}_{(L-1)\rightarrow 2}(1,\ldots (L-1))$ acts, the site originally labeled $L$ becomes site $3$.}. This symmetric choice matches our convention of using parity even gauge transformations.  

To simplify calculations later,  it will be useful to decompose the $\dot{\gen{S}}$ commutator according to $\gen{B}$ charge as described in (\ref{eq:bchargelabel}). Dropping gauge terms, we then have
\[ \label{eq:sdotcommutatorbybcharge}
\comm{\dot{\gen{S}}}{g^{-2}\, \delta\gen{D}}_{L \rightarrow 2} = \sum_{k=0}^{L-2}  \comm{\dot{\gen{S}}^{[0]}_{1 \rightarrow 1}}{\gen{D}^{[k]}_{L \rightarrow 2}} + \sum_{k=0}^{L-3} \comm{\dot{\gen{S}}^{[0]}_{2 \rightarrow 1}}{\gen{D}^{[k]}_{(L-1) \rightarrow 2}} =0.
\]

We can replace $\dot{\gen{S}}$ and $\dot{\gen{s}}$ in (\ref{eq:sdeltadcommutator}) with $\gen{S}$ and $\gen{s}$ to obtain the analogous equations for $\gen{S}$. Similarly for $\gen{K}$ we have 
\< \label{eq:briefkcommutator}
\comm{\gen{K}}{g^{-2}\, \delta\gen{D}}_{L \rightarrow 2} \eq  \delta_{L3} \gen{k}_{3 \rightarrow 2} +  \delta_{L4} \gen{k}_{4 \rightarrow 2} \\
\eq \comm{\gen{K}_{1 \rightarrow 1}}{\gen{D}_{L \rightarrow 2}} + \delta_{L\neq2} \, \comm{\gen{K}_{2 \rightarrow 1}}{\gen{D}_{(L-1) \rightarrow 2}} + \theta(L-4) \, \comm{\gen{K}_{3 \rightarrow 1}}{\gen{D}_{(L-2) \rightarrow 2}}. \notag
\>
(\ref{eq:fullkcommutator}) gives the lengthy local expression for this commutator.

We finish the section by defining, for later convenience,
 \[ \label{eq:definem}
 \gen{M}_{L \rightarrow 2}= \comm{(\gen{K}_{11})_{1 \rightarrow 1}}{\gen{D}_{L \rightarrow 2}} .
 \]
 Note that $\gen{M}$ has the same quantum numbers as $\gen{K}_{11}$. Then, like $\gen{D}_{L \rightarrow 2}$, $\gen{M}$ can be split into components according to $B$-charge. Using (\ref{eq:briefkcommutator}), $\gen{M}^{[k]}_{L \rightarrow 2}$ for $0 \leq k \leq L-2$ can be written as
 \<
 \gen{M}^{[0]}_{2 \rightarrow 2} \eq 0, \label{eq:mhelicitycomponents} \\
\gen{M}^{[k]}_{3 \rightarrow 2} \eq - \comm{(\gen{K}^{[0]}_{11})_{2 \rightarrow 1}}{\gen{D}^{[k]}_{2 \rightarrow 2}} - \comm{(\gen{K}^{[1]}_{11})_{2 \rightarrow 1}}{\gen{D}^{[k-1]}_{2\rightarrow 2}} + (\gen{k}^{[k]}_{11})_{3 \rightarrow 2},  \nln
\gen{M}^{[k]}_{4 \rightarrow 2} \eq - \comm{(\gen{K}^{[0]}_{11})_{2 \rightarrow 1}}{\gen{D}^{[k]}_{3 \rightarrow 2}} - \comm{(\gen{K}^{[1]}_{11})_{2 \rightarrow 1}}{\gen{D}^{[k-1]}_{3\rightarrow 2}}  - \comm{(\gen{K}^{[1]}_{11})_{3 \rightarrow 1}}{\gen{D}^{[k-1]}_{2 \rightarrow 2}}  + \delta_{k1}(\gen{k}^{[k]}_{11})_{4 \rightarrow 2},
   \nln
 \gen{M}^{[k]}_{L \rightarrow 2} \eq- \comm{(\gen{K}^{[0]}_{11})_{2 \rightarrow 1}}{\gen{D}^{[k]}_{(L-1) \rightarrow 2}} - \comm{(\gen{K}^{[1]}_{11})_{2 \rightarrow 1}}{\gen{D}^{[k-1]}_{(L-1) \rightarrow 2}} - \comm{(\gen{K}^{[1]}_{11})_{3 \rightarrow 1}}{\gen{D}^{[k-1]}_{(L-2) \rightarrow 2}}  \nl
  \quad (L > 4). \notag
 \>
The nonvanishing components of $\gen{k}$ are $\gen{k}^{[0]}_{3 \rightarrow2}$, $\gen{k}^{[1]}_{3 \rightarrow 2}$, and $\gen{k}^{[1]}_{4 \rightarrow 2}$.

\section{The unique leading order solution for $\delta\gen{D}$ \label{sec:theconstruction}}

We begin this section by giving a brief argument that $\gen{D}_{L \rightarrow 2}$ is fixed by superconformal symmetry. Since the one-loop dilatation generator $\gen{D}_{2 \rightarrow 2}$ is fixed by superconformal symmetry, we can assume $\gen{D}_{L' \rightarrow 2}$ is fixed by superconformal symmetry for $L' < L$. Also, for some given $\vec{n}_1, \vec{n}_2$, assume that we know 
\[
\langle \vec{n}_1 \vec{n}_2|\gen{D}_{L \rightarrow 2},
\]
that is we know all matrix elements $\langle \vec{n}_1 \vec{n}_2|\gen{D}_{L \rightarrow 2} \state{\vec{m}_1 \ldots \vec{m}_L}$. These assumptions imply that superconformal symmetry fixes
\[ \label{eq:inductivestep}
\langle \vec{n}_1 \vec{n}_2|\gen{J}_{1 \rightarrow 1} \gen{D}_{L \rightarrow 2},
\]
for any $\alg{psu}(2,2|4)$ generator $\gen{J}$. This is because $ \comm{\gen{J}_{1 \rightarrow 1}}{ \gen{D}_{L \rightarrow 2}}$, if nonzero, is given in terms of the fixed    $\gen{D}_{L' \rightarrow 2}$ for $L' < L$, and the known $\gen{J}_{2 \rightarrow 1}$ and $\gen{J}_{3 \rightarrow 1}$ (and possibly generalized gauge transformations). 

We will now complete the argument by induction. Consider $\vec{n}_1,\, \vec{n_2}$ in the bosonic $\alg{sl}(2)$ sector with
\[
 \vec{n}_1=(n_1, 0, n_1, 0, 1, 1, 0, 0), \quad \vec{n}_2=(n_2, 0, n_2, 0, 1, 1, 0, 0).
 \]
 Then $\langle \vec{n}_1 \vec{n}_2|\gen{D}_{L \rightarrow 2}$ for $L > 2$ vanishes (so of course it is known)  because $\gen{D}_{L \rightarrow 2}$ must insert at least one $\mathbf{a}^{2\dagger}$ or $\mathbf{b}^{2\dagger}$ \footnote{This is an explanation of the well-known fact that length is conserved in the $\alg{sl}(2)$ sector.}. As used in \cite{Beisert:2003jj}, every irreducible representation in the tensor product of two field strength multiplets has an element in the bosonic $\alg{sl}(2)$ sector. Therefore, by acting with linear combinations of \emph{all} $\alg{psu}(2,2|4)$ $\gen{J}_{1 \rightarrow 1}$ on $\alg{sl}(2)$ sector $\langle \vec{n}_1 \vec{n}_2|$ we can reach any $\langle \vec{n}_1 \vec{n}_2|$ of the full theory. Repeatedly applying the inductive step of (\ref{eq:inductivestep}), it follows that all $\langle \vec{n}_1 \vec{n}_2|\gen{D}_{L \rightarrow 2}$ are fixed by superconformal symmetry.

The remainder of this section describes a construction of  all $\gen{D}_{L \rightarrow 2}$. In addition to confirming the argument given above, we will use this construction later to test the proposed relation to scattering amplitudes. The following chapters are largely independent of this technical construction.  

The construction starts from the known $\gen{D}_{2 \rightarrow 2}$, the classical $1 \rightarrow 1$ interactions, the known length-changing interactions of $\gen{K}$, and the generalized gauge transformations $\gen{k}$.  We first give a brief summary of the construction.  Below we will give more details, with an explanation of the key identities left for Section \ref{sec:deriveidentities}.   Since we know $\gen{D}_{2 \rightarrow 2}$, we will compute  $\langle\vec{n}_1\vec{n}_2|\gen{D}^{[k]}_{L \rightarrow 2}|\vec{m}_1 \ldots \vec{m}_L\rangle$, assuming $\gen{D}^{[k]}_{L' \rightarrow 2}$ is known for $L'<L$.  Using $\gen{K}_{1\rightarrow1}$ and $\gen{M}$ (\ref{eq:definem}), we can then construct $\langle\vec{n}_1\vec{n}_2|\gen{D}^{[k]}_{L \rightarrow 2}|\vec{m}_1 \ldots \vec{m}_L\rangle$ in terms of known quantities, provided $(k$, $L$, $\vec{n}_1$, $\vec{n}_2)$   satisfy certain properties. Next we  use the vanishing commutator between $\gen{D}^{[k]}_{L \rightarrow 2}$ and the Lorentz generators and  $\gen{Q}$ to obtain $\langle\vec{n}_1\vec{n}_2|\gen{D}^{[k]}_{L \rightarrow 2}|\vec{m}_1 \ldots \vec{m}_L\rangle$ for additional classes of matrix elements.  The remaining matrix elements follow from spin chain parity or the conjugation symmetry (\ref{eq:conjugation}). The construction can then be repeated for $L+1$. We use parity or conjugation symmetry to shorten the construction, but as the argument above shows, these symmetries are not necessary for fixing $\gen{D}_{L \rightarrow 2}$.

\subsection{Construction of $\gen{D}_{L \rightarrow 2}$}

Following (\ref{eq:parity}) we defined $n_{i,f}$ as the number of fermionic oscillators on site $i$, so we have
\[
n_{i, f} = \sum_{\alpha=5}^8 n_{i,\alpha}.
\]
Consider $\langle\vec{n}_1\vec{n}_2|\gen{D}^{[k]}_{L \rightarrow 2}|\vec{m}_1 \ldots \vec{m}_L\rangle$  when the following four conditions hold:
\begin{enumerate}
\item $n_{2,2}=n_{2,4}=0$
\item $n_{2,f} \geq 2$
\item $n_{1,f} \leq n_{2,f}$
\item $k \leq \half (L-2)$
\end{enumerate}
Then the matrix elements follow from known quantities as\footnote{This identity holds more generally. In particular, for $L > 4$ or for $L=4,k=1$, the only other conditions we need are that $n_{2,2}=n_{2,4}=0$.},
\[
\langle\vec{n}_1\vec{n}_2|\gen{D}^{[k]}_{L \rightarrow 2}|\vec{m}_1 \ldots \vec{m}_L\rangle  =\sum_{j=0}^{\min(n_{2,1},n_{2,3})-1} \sum_{i=0}^{j} r_{i,j} \langle\vec{n}'_{1,ij}\vec{n}'_{2,j}|\gen{M}^{[k]}_{L \rightarrow 2}((\gen{K}_{11})_{1 \rightarrow 1})^i|\vec{m}_1 \ldots \vec{m}_L\rangle,
\]
where 
\<
r_{i,j} \eq (-1)^{j-i}\binom{j}{i}\frac{(n_{1,1}+j-i)!\, (n_{1,3}+j-i)! \, (n_{2,1}-j-1)! \, (n_{2,3}-j-1)!}{n_{1,1}!\, n_{1,3}!\, n_{2,1}!\, n_{2,3}!}, \notag \\
\vec{n}'_{1,ij} \eq \vec{n}_1 +(j-i) \vec{\delta}_{1,3}, \quad  \vec{n}'_{2,j} = \vec{n}_2 -(j+1)\, \vec{\delta}_{1,3}, \label{eq:kidentity}
\>
Recall that we wrote an expression for $\gen{M}^{[k]}_{L \rightarrow 2}$  in (\ref{eq:mhelicitycomponents}), which shows that $\gen{M}^{[k]}_{L \rightarrow 2}$ follows from known quantities including $\gen{D}_{(L-1) \rightarrow 2}$. A derivation of this identity (\ref{eq:kidentity}) will be given below in Section \ref{sec:deriveidentities}.

Matrix elements not satisfying condition 1 (but satisfying conditions 2-4), that is matrix elements for arbitrary $n_{2,2}$ and $n_{2,4}$,  are then given by the identity\footnote{This identity is always valid.}
\begin{gather} \label{eq:lldotidentity}
\langle\vec{n}_1\vec{n}_2|\gen{D}_{L \rightarrow 2}|\vec{m}_1 \ldots \vec{m}_L\rangle  =  \sum_{i=0}^{n_{2,2}}\sum_{j=0}^{n_{2,4}}r_{ij} \langle \vec{n}_{ij} \vec{n}^{0} |\gen{D}_{L \rightarrow 2} (\gen{L}^1_2)^{i}(\dot{\gen{L}}^1_2)^{j}|\vec{m}_1 \ldots \vec{m}_L\rangle ,
\\
r_{ij}  = (-1)^{n_{2,2}-i}(-1)^{n_{2,4}-j}\binom{n_{2,2}}{i}\binom{n_{2,4}}{j}\frac{(n_{1,2}+n_{2,2}-i)! }{n_{1,2}!\,n_{2,2}!}\frac{(n_{1,4}+n_{2,4}-j)! }{n_{1,4}!\,n_{2,4}!} , 
\notag
\\
\vec{n}^{0}= \vec{n}_2 - n_{2,2}(\vec{\delta}_2-\vec{\delta}_1) - n_{2,4}(\vec{\delta}_4-\vec{\delta}_3)= (n_{2,1}+n_{2,2}, 0, n_{2,3}+n_{2,4}, 0, n_{2,5}, n_{2,6}, n_{2,7}, n_{2,8}), \notag \\
 \vec{n}_{ij} = \vec{n}_1+ (n_{2,2}-i) (\vec{\delta}_2- \vec{\delta}_1) + (n_{2,4}-j) (\vec{\delta}_4- \vec{\delta}_3).  \notag 
\end{gather}
Note that this identity is trivially satisfied when $n_{2,2}=n_{2,4}=0$, since then the summations only include the $i=j=0$ term, $\vec{n}^0=\vec{n_2}$, and $\vec{n}_{ij}=\vec{n}_1$. (\ref{eq:lldotidentity}) follows straightforwardly from Lorentz symmetry, as we will show below  in Section \ref{sec:deriveidentities}. The key point  is that this gives matrix element of  $\gen{D}^{[k]}_{L \rightarrow 2}$ satisfying conditions 2-4 in terms of matrix elements satisfying all four conditions (which are known from (\ref{eq:kidentity})), since the second final site $\vec{n}^{0}$ has $n^{0}_2=n^{0}_4=0$.  At this point, we have expressions for general matrix elements satisfying conditions 2-4.

Next consider matrix elements satisfying only conditions 3 and 4, but not satisfying condition 2. This means that $n_{2,f} < 2$ and $n_{1,f} \leq n_{2,f}$. Therefore, there  are two  types of fermionic oscillators that are not excited for the final state $\langle \vec{n}_1 \vec{n}_2|$. We label these fermionic oscillators\footnote{When there are more than two that are not excited, which two we choose does not matter.} $\mathbf{d}^{\dagger A}$ and $\mathbf{d}^{\dagger B}$. Then we use the identity valid for $A < B$
\<
\langle\vec{n}_1\vec{n}_2|\gen{D}^{[k]}_{L \rightarrow 2}|\vec{m}_1  \ldots \vec{m}_L\rangle \eq  (-1)^{\sigma_{AB}}\langle\vec{n}_1(\vec{n}_2+\vec{\delta}_{1(A+4)} +\vec{\delta}_{1(B+4)} )|\gen{D}^{[k]}_{L \rightarrow 2}\gen{Q}^{1A}\gen{Q}^{1B}|\vec{m}_1 \ldots \vec{m}_L\rangle, \label{eq:qidentity} \notag \\
\quad \sigma_{AB} \eq \sum_{\alpha=A+5}^{B+3}  n_{2,\alpha},
\>
to write this state with $n_{2,f} < 2$ in terms of states which do satisfy condition 2 ($n_{2,f} \geq 2$) and therefore are already known. This third identity will also be derived below.

This construction now applies for arbitrary matrix elements satisfying conditions 3 and 4. The remaining matrix elements not satisfying one or both of these conditions, follow from parity (\ref{eq:parity}) and conjugation symmetry (\ref{eq:conjugation}) as:
\<
\langle\vec{n}_1\vec{n}_2|\gen{D}^{[k]}_{L \rightarrow 2}|\vec{m}_1  \ldots \vec{m}_L\rangle \eq (-1)^{L + (n_{1,f})(n_{2,f}) + \prod_{i<j}( m_{i,f})(m_{j,f})}  \langle\vec{n}_2\vec{n}_1|\gen{D}^{[k]}_{L \rightarrow 2}|\vec{m}_L \vec{m}_{L-1}  \ldots \vec{m}_1\rangle \nln
\langle\vec{n}_1\vec{n}_2|\gen{D}^{[k]}_{L \rightarrow 2}|\vec{m}_1  \ldots \vec{m}_L\rangle \eq \langle\vec{n}^\ast_1\vec{n}^\ast_2|\gen{D}^{[L-2-k]}_{L \rightarrow 2}|\vec{m}^\ast_1  \ldots \vec{m}^\ast_L\rangle. \label{eq:parityandconjugation}
\>
We now have expressions for all $\langle\vec{n}_1\vec{n}_2|\gen{D}^{[k]}_{L \rightarrow 2}|\vec{m}_1  \ldots \vec{m}_L\rangle$, and can repeat this for $(L+1)$. Therefore, this construction gives the unique solution for $\gen{D}_{L \rightarrow 2}$ for all $L$. 

The above construction  may exaggerate the importance of supersymmetry for fixing $\gen{D}_{L \rightarrow 2}$. For $L >4$, one can construct $\gen{D}_{L \rightarrow 2}$ simply using the $\gen{K}$ identity (\ref{eq:kidentity})  and the Lorentz symmetry generators identity (\ref{eq:lldotidentity}), without needing to use the supersymmetry generators\footnote{It is intriguing that \cite{Braun:2009vc} found $2 \rightarrow 3$ BFKL kernels for QCD just using conformal symmetry, Lorentz transformations and previously known results.}.

Using the symmetry relations satisfied by $\delta \gen{D}$ and  $\gen{M}$, it is possible to significantly simplify these expressions. However, we will not pursue that in this work, since relating $\gen{D}_{L \rightarrow 2}$ to scattering amplitudes will immediately give us a simpler and more elegant expression for $\gen{D}_{L \rightarrow 2}$.

This construction for $\gen{D}_{L \rightarrow 2}$ then also gives $\gen{D}_{2 \rightarrow L}$  via the Hermitian transformation rule (\ref{eq:hermitianconjugationofd}).

\subsection{Derivation of identities \label{sec:deriveidentities}}
We now give brief derivations of the three identities (\ref{eq:kidentity}), (\ref{eq:lldotidentity}), and (\ref{eq:qidentity}), used for the above construction. We start with an explanation for (\ref{eq:kidentity}). From (\ref{eq:definepsu224oscillator}) and  (\ref{eq:mhelicitycomponents}), we have
\[
\langle\vec{n}|(\gen{K}_{11})_{1 \rightarrow 1} = \langle\vec{n}+\vec{\delta}_{13}|(n_1+1)(n_3+1), \quad \comm{(\gen{K}_{11})_{1 \rightarrow 1}}{\gen{D}^{[k]}_{L \rightarrow 2}} = \gen{M}^{[k]}_{L \rightarrow 2}.
\]
It follows that
\<
\langle\vec{n}_1(\vec{n}_2+ \vec{\delta}_{13})|\gen{D}^{[k]}_{L \rightarrow 2} \eq  
\frac{1}{(n_{2,1}+1)(n_{2,3}+1)} \bigg(\langle\vec{n}_1\vec{n}_2|\gen{D}^{[k]}_{L \rightarrow 2}(\gen{K}_{11})_{1 \rightarrow 1}
+ \langle\vec{n}_1\vec{n}_2| \gen{M}^{[k]}_{L \rightarrow 2}
\notag   \\
& &  - (n_{1,1}+1)(n_{1,3}+1) \langle(\vec{n}_1+\vec{\delta}_{13})\vec{n}_2|\gen{D}^{[k]}_{L \rightarrow 2} \bigg). \label{eq:kidentitykernel}
\>
Next define an excitation vector $\vec{n}^{000}$ that has at least three of its first four entries equal to 0 (assuming condition 1, $n_{2,2}=n_{2,4}=0$), 
\[
\vec{n}^{000}=\vec{n}_2- \min(n_{2,1}, n_{2,3})\vec{\delta}_{13}.
\]
 Applying the identity (\ref{eq:kidentitykernel}) $n_{\min} =\min(n_{2,1}, n_{2,3})$ times,   keeping track of combinatoric factors, yields
\begin{gather}
\langle\vec{n}_1\vec{n}_2|\gen{D}^{[k]}_{L \rightarrow 2}|\vec{m}_1 \ldots \vec{m}_L\rangle  = \sum_{j=0}^{n_{\min}-1} \sum_{i=0}^{j} r_{i,j} \langle\vec{n}'_{1,ij}\vec{n}'_{2,j}|\gen{M}^{[k]}_{L \rightarrow 2}((\gen{K}_{11})_{1 \rightarrow 1})^i|\vec{m}_1 \ldots \vec{m}_L\rangle
\\
+  \sum_{i=0}^{n_{\mathrm{min}}} r_{i,n_{\mathrm{min}}} \langle\vec{n}'_{1,i \, n_{\mathrm{min}}}\vec{n}^{000}|\gen{D}^{[k]}_{L \rightarrow 2}((\gen{K}_{11})_{1 \rightarrow 1})^i|\vec{m}_1 \ldots \vec{m}_L\rangle,  \notag
\end{gather}
where
\begin{gather}
r_{i,j} =(-1)^{j-i}\binom{j}{i}\frac{(n_{1,1}+j-i)!\, (n_{1,3}+j-i)! \, (n_{2,1}-j- \delta_{j \neq n_{\mathrm{min}}} )! \, (n_{2,3}-j-\delta_{j\neq n_{\mathrm{min}}})!}{n_{1,1}!\, n_{1,3}!\, n_{2,1}!\, n_{2,3}!}, \notag \\
\vec{n}'_{1,ij} = \vec{n}_1 +(j-i) \vec{\delta}_{1,3}, \quad  \vec{n}'_{2,j} = \vec{n}_2 -(j+1)\, \vec{\delta}_{1,3},
\end{gather}
and $\delta_{j\neq n_{\mathrm{min}}} = (1 - \delta_{j\, n_{\mathrm{min}}})$. This identity holds for general k and general $L > 2$. The key observation is that when the four conditions hold, the second line vanishes as we now explain. Since $k < \half (L-2)$ (for $L > 2$), $\gen{D}^{[k]}_{L \rightarrow 2}$ inserts at least one singlet pair of $\mathbf{b}^\dagger$ oscillators. On the other hand,  since we assume $n_{2,f} \geq 2$ and due to the central charge constraint, the excitations on the second site ($\vec{n}^{000}$) include no $\mathbf{b}^\dagger$  excitations.  Therefore, the matrix element involving $\gen{D}^{[k]}_{L \rightarrow 2}$ on the second line must vanish as stated above.  Dropping the second line, we recover precisely the identity (\ref{eq:kidentity}).

Next we explain the origin of the Lorentz generators identity (\ref{eq:lldotidentity}). Using the vanishing commutator of $\gen{L}^1_2$ with $\gen{D}_{L \rightarrow 2}$, and the action of $\gen{L}^1_2$ (\ref{eq:definepsu224oscillator})
\[
\langle \vec{n} |\gen{L}^1_2 = \langle \vec{n}+\vec{\delta}_2-\vec{\delta}_1 |(n_2+1),
\]
we have the identity
\<
\lefteqn{\langle\vec{n}_1(\vec{n}_2+ \vec{\delta}_2-\vec{\delta}_1)|\gen{D}_{L \rightarrow 2} =} 
\\
& & \frac{1}{n_{2,2}+1} \bigg( \langle\vec{n}_1\vec{n}_2|\gen{D}_{L \rightarrow 2}\gen{L}^1_2  - (n_{1,2}+1) \langle(\vec{n}_1+ \vec{\delta}_2-\vec{\delta}_1 )\vec{n}_2|\gen{D}_{L \rightarrow 2} \bigg). \notag
\>
Applying this relation $n_{2,2}$ times starting from a vector with second entry zero, $\vec{n}^{02}=\vec{n}_2 - n_{2,2} (\vec{\delta}_2 - \vec{\delta}_1)$, and acting on a general $L$-site state, we obtain
\begin{gather}
\langle\vec{n}_1\vec{n}_2|\gen{D}_{L \rightarrow 2}|\vec{m}_1 \ldots \vec{m}_L\rangle  =  \sum_{i=0}^{n_{2,2}}r_{2,i} \langle \vec{n}_i' \vec{n}^{02} |\gen{D}_{L \rightarrow 2} (\gen{L}^1_2)^{i}|\vec{m}_1 \ldots \vec{m}_L\rangle , \notag
\\
r_{2,i}  = (-1)^{n_{2,2}-i}\binom{n_{2,2}}{i}\frac{(n_{1,2}+n_{2,2}-i)! }{n_{1,2}!\,n_{2,2}!} , \quad \vec{n}_i' = \vec{n}_1+ (n_{2,2}-i) (\vec{\delta}_2- \vec{\delta}_1). \label{eq:lrecursion}
\end{gather}
Repeating the analogous steps with $\dot{\gen{L}}^1_2$ instead, and defining a vector with fourth entry zero as $\vec{n}^{04}=\vec{n}_2 - n_{2,4} (\vec{\delta}_4 - \vec{\delta}_3)$, we have
\begin{gather}
\langle\vec{n}_1\vec{n}_2|\gen{D}_{L \rightarrow 2}|\vec{m}_1 \ldots \vec{m}_L\rangle  =  \sum_{i=0}^{n_{2,4}}r_{4,i} \langle \vec{n}_i' \vec{n}^{04} |\gen{D}_{L \rightarrow 2} (\dot{\gen{L}}^1_2)^{i}|\vec{m}_1 \ldots \vec{m}_L\rangle , \notag
\\
r_{4,i}  = (-1)^{n_{2,4}-i}\binom{n_{2,4}}{i}\frac{(n_{1,4}+n_{2,4}-i)! }{n_{1,4}!\,n_{2,4}!} , \quad \vec{n}_i' = \vec{n}_1+ (n_{2,4}-i) (\vec{\delta}_4- \vec{\delta}_3). \label{eq:ldotrecursion}
\end{gather}
The identity (\ref{eq:lldotidentity}) then follows straightforwardly from combining (\ref{eq:lrecursion} - \ref{eq:ldotrecursion}).

Finally, for the identity involving $\gen{Q}$ (\ref{eq:qidentity}), we start from
\[
\langle \vec{n}|\gen{Q}^{1A} = \langle \vec{n}+ \vec{\delta}_{1(A+4)}| (-1)^{\sum_{\alpha=5}^{A+3}  n_\alpha}\delta_{n_{A+4},0}.
\]
Then, assuming $A < B$ and $0=n_{1,(A+4)}=n_{2,(A+4)}=n_{1,(B+4)}=n_{2,(B+4)}$, it follows that

\begin{gather} 
\langle\vec{n}_1\vec{n}_2|\gen{D}^{[k]}_{L \rightarrow 2}|\vec{m}_1 \ldots \vec{m}_L\rangle  =  (-1)^{\sigma_{AB}} \langle\vec{n}_1(\vec{n}_2+\vec{\delta}_{1(A+4)} +\vec{\delta}_{1(B+4)})|\gen{Q}^{1A}\gen{Q}^{1B}\gen{D}^{[k]}_{L \rightarrow 2}|\vec{m}_1 \ldots \vec{m}_L\rangle  \\
= (-1)^{\sigma_{AB}} \langle\vec{n}_1(\vec{n}_2+\vec{\delta}_{1(A+4)} +\vec{\delta}_{1(B+4)} )|\gen{D}^{[k]}_{L \rightarrow 2}\gen{Q}^{1A}\gen{Q}^{1B}|\vec{m}_1\ldots \vec{m}_L\rangle, \quad \sigma_{AB} = \sum_{\alpha=A+5}^{B+3}  n_{2,\alpha}. \notag
\end{gather}
The second equality follows from the vanishing commutator of $\gen{Q}$ and $\gen{D}_{L \rightarrow 2}$. This is precisely the identity (\ref{eq:qidentity}) we used in the construction, completing our derivation of the necessary identities.

\section{Translating to spinor helicity superspace \label{sec:translation}}

Anticipating the next section, we translate between the oscillator notation and a spinor helicity superspace notation as\footnote{This is natural since the oscillators and the spinor-helicity variables satisfy the same commutation relations. For example, 
\[
\comm{\mathbf{a}_{\alpha}}{\mathbf{a}^{\dagger \beta}}  = \delta^\beta_\alpha = \comm{\partial_\alpha}{\lambda^\beta}. \notag
\]
}
\[
\begin{array}{r@{\leftrightarrow}lr@{\leftrightarrow}lr@{\leftrightarrow}l}
\mathbf{a}^{\dagger \alpha}_i & \lambda^\alpha_i, & \mathbf{b}_i^{\dagger \dot{\beta}} & \bar{\lambda}^{\dot \beta}_i, & \mathbf{d}^{\dagger C}_i  & \eta^C_i,
 \\
\mathbf{a}_{i,\alpha} & \partial_{i,\alpha}, & \mathbf{b}_{i,\dot \beta} & \bar{\partial}_{i,\dot \beta}, & \mathbf{d}_{i,C}  & \partial_{i,C}. \label{eq:osctranslation}
\end{array}
\]
The translation is to complex conjugate pairs of spinors $(\lambda, \bar{\lambda})$ since we will only consider $(3,1)$-signature amplitudes. We will use the convenient abbreviations of
\<
\Lambda_i \eq(\lambda^\alpha_i,  \bar{\lambda}^{\dot \beta}_i,\eta^C_i),
\nln
\mathrm{d} \Lambda_i \eq \mathrm{d}^2 \lambda^1_i \, \mathrm{d}^2 \lambda^2_i \,  \mathrm{d}^4 \eta_i.
\>
Also, $\mathbf{c}$ oscillators map to $\bar{\eta}$ variables, where  the $\bar{\eta}$ are the Grassmann Fourier transform of $\eta$,
\[
\begin{array}{r@{\leftrightarrow}lr@{\leftrightarrow}l}
\mathbf{c}^\dagger_{i,A}  & \bar{\eta}_{i,A} & \mathbf{c}^A_i  & \bar{\partial}^A_i.
\end{array} \label{eq:introduceetabar}
\]
We introduce spinor helicity superspace states $\state{\Lambda}$, and we define their inner product as
\[
\langle \Lambda' | 
\Lambda \rangle = \delta^{4|4}(\Lambda-\Lambda'). \label{eq:Lambdainnerproduct}
\]
We also write tensor products of these states as $\state{\Lambda_1 \ldots \Lambda_L}$. Under the translation a generic length-$L$ state of the spin chain becomes a polynomial state $\state{\mathcal{P}_L}$ in  spinor-helicity variables as 
 \<
 \state{\vec{n}_1 \ldots \vec{n}_L} & \leftrightarrow & \state{\mathcal{P}_L}= \int \mathrm{d}\Lambda_1 \ldots \mathrm{d}\Lambda_L \, \mathcal{P}_L(\Lambda_1,\ldots\Lambda_L) \state{\Lambda_1 \ldots \Lambda_L},
 \nln
 \mathcal{P}(\Lambda_1, \ldots \Lambda_L)  \eq  \prod_{i=1}^L \textstyle \prod  \limits_{\alpha=1}^2  (\lambda_{i}^\alpha)^{n_{i,\alpha}} \textstyle \prod  \limits_{\dot{\beta}=1}^2  (\bar{\lambda}^{\dot{\beta}}_i)^{n_{i,2+\dot{\beta}}} \textstyle \prod  \limits_{C=1}^4  (\eta^C_i)^{n_{i,4+C}}.
 \>
 Note that the second line realizes the translation (\ref{eq:osctranslation}), and the first line combined with (\ref{eq:Lambdainnerproduct}) implies $\langle \Lambda_1 \ldots \Lambda_L | \mathcal{P}_L \rangle = \mathcal{P}(\Lambda_1, \ldots \Lambda_L)$.  The central charge constraint (\ref{eq:centralchargecondition}) implies that any $\mathcal{P}$ transforms homogeneously under a phase shift applied to any individual argument,
 \[
 \mathcal{P}(\Lambda_1, \ldots e^{i \phi}  \Lambda_i, \ldots \Lambda_L) = e^{-2 i \phi}  \mathcal{P}( \Lambda_1, \ldots \Lambda_i, \ldots \Lambda_L), \quad e^{i \phi} \Lambda = (e^{i \phi} \lambda, e^{-i \phi} \bar{\lambda}, e^{-i \phi} \eta). \label{eq:phaseshift}
 \]
From  (\ref{eq:Lambdainnerproduct}), we see that the identity operator $\mathbf{1}$ in the $\Lambda$-basis is  
\[
\mathbf{1} = \sum_{L'=1}^\infty \prod_{i=1}^{L'} \int \mathrm{d} \Lambda_i  \state{\Lambda_i} \langle \Lambda_i|, \label{eq:lambdabasisidentity}
\]
though when working with a $\mathcal{P}_L$ of fixed number of arguments, we only need one term of the infinite sum in (\ref{eq:lambdabasisidentity}).

For the classical linear generators, we find
\[
\begin{array}{r@{=}lr@{=}lr@{=}l}
\gen{L}^\alpha_\beta & \lambda^\alpha \partial_\beta -\half \delta^\alpha_\beta \lambda^\gamma \partial_\gamma, & \gen{Q}^{\alpha B}&\lambda^\alpha \eta^B , & \gen{P}^{\alpha \dot{\beta}} & \lambda^\alpha \bar{\lambda}^{\dot{\beta}} ,
\\
 \dot{\gen{L}}^{\dot{\alpha}}_{\dot{\beta}} & \bar{\lambda}^{\dot{\alpha}} \bar{\partial}_{\dot{\beta}} -\half \delta^{\dot{\alpha}}_{\dot{\beta}} \bar{\lambda}^{\dot{\gamma}} \bar{\partial}_{\dot{\gamma}} , &   \gen{S}_{\alpha B}&\partial_\alpha \partial_B ,& \gen{K}_{\alpha \dot{\beta}} & \partial_\alpha \bar \partial_{\dot{\beta}}, 
\\
\gen{R}_A^B & \eta^B\partial_A -\sfrac{1}{4} \delta_A^B \eta^B \partial_A ,&  \dot{\gen{Q}}^{\dot{\alpha}}_B&\bar{\lambda}^{\dot{\alpha}} \bar \partial_B, & \gen{C}\, \, \, &\half \lambda^\gamma \partial_\gamma - \half \bar{\lambda}^{\dot{\gamma}} \bar \partial_{\dot{\gamma}}-\eta^C\partial_C + 1 ,
\\
\gen{D}& \half \lambda^\gamma \partial_\gamma + \half \bar{\lambda}^{\dot{\gamma}} \bar \partial_{\dot{\gamma}}+ 1,  &
  \dot{\gen{S}}_{\dot{\alpha}}^B&\bar \partial_{\dot{\alpha}} \eta^ B, &   \gen{B} \, \,  &  \eta^C\partial_C.  \label{eq:classicalgeneratorinsuperspace}
\end{array}
\]
When these generators act on a polynomial state, we sum the actions of the differential operators on each $\Lambda_i$ argument as usual. 
 
 Using (\ref{eq:Hisrotatingoscillators}), it is straightforward to transform the expressions (\ref{eq:defineinteractingsupercharge})  for $\gen{S}_{2 \rightarrow 1}$ and $\dot{\gen{S}}_{2 \rightarrow 1}$ : 
 \<
 \langle \bar \Lambda | (\gen{S}_{\alpha C})_{2 \rightarrow 1} \state{\mathcal{P}_2} \eq \varepsilon_{\alpha \beta} \lambda^\beta \int \mathrm{d}^4\bar{\eta}' \bar{\eta}'_C \int_0^{\pi/2} \mathrm{d} \theta \, \mathcal{P}(\bar \Lambda_1, \bar \Lambda_2),
 \nln
 \langle \Lambda | (\dot{\gen{S}}_{\dot{\alpha}}^C)_{2 \rightarrow 1} \state{\mathcal{P}_2} \eq \varepsilon_{\dot{\alpha}\dot{ \beta}} \bar{\lambda}^{\dot{\beta}} \int \mathrm{d}^4\eta' \eta'^C \int_0^{\pi/2} \mathrm{d} \theta \, \mathcal{P}(\Lambda_1, \Lambda_2),
 \nln
 \lambda^\alpha_1 \eq \lambda^\alpha \cos \theta, \quad \lambda^\alpha_2 = \lambda^\alpha \sin \theta,
 \nln
 \bar{\lambda}^{\dot{\alpha}}_1 \eq \bar{\lambda}^{\dot{\alpha}} \cos \theta, \quad \bar{\lambda}^{\dot{\alpha}}_2 = \bar{\lambda}^{\dot{\alpha}} \sin \theta,
 \nln
 \bar{\eta}_{1,A} \eq \bar{\eta}_A \cos \theta - \bar{\eta}'_A \sin \theta, \quad \bar{\eta}_{2,A} = \bar{\eta}_A \sin \theta + \bar{\eta}'_A \cos \theta,
 \nln
 \eta^A_1 \eq \eta^A \cos \theta - \eta'^A \sin \theta, \quad \eta^A_2 = \eta^A \sin \theta + \eta'^A \cos \theta. \label{eq:sandsdotinspinorhelicity}
 \>
 In the first line expression for $(\gen{S}_{a C})_{2 \rightarrow 1}$,  we introduced $\bar \Lambda$ which uses $\bar \eta$ instead of $\eta$. For both $\gen{S}$ and $\dot{\gen{S}}$, since we only consider $\mathcal{P}$ that transform homogeneously under the phase shift for each argument, it is straightforward to check that the resulting polynomial in $\Lambda$ also transforms homogeneously, and we do not need additional phase integrations. 
 
 Finally, in this notation the $2 \rightarrow 2$ dilatation generator interactions (\ref{eq:regularizedd2}) become
 \[
  \langle \Lambda _1 \Lambda_2|\gen{D}_{2 \rightarrow 2} \state{\mathcal{P}} = \frac{4 \, }{2 \pi i} \int_{0}^{2 \pi} \mathrm{d} \phi \int_0^{\pi /2} \mathrm{d} \theta \,  \cot \theta \bigg(\mathcal{P}(\Lambda_1,\Lambda_2) - e^{2 i \phi}\mathcal{P}(\Lambda'_1, \Lambda'_2) \bigg). \label{eq:d2insuperspace}
 \]
Now the change of variables is (suppressing $\alg{su}(2)$ and $\alg{su}(4)$ indices)
 \begin{align}
 \lambda'_1 & = e^{i \phi}(\lambda_1 \cos \theta + \lambda_2 \sin \theta ), &  \lambda'_2 & = -\lambda_1 \sin \theta + \lambda_2 \cos \theta , \notag \\
 \bar{\lambda}'_1 & = e^{-i \phi}(\bar{\lambda}_1 \cos \theta + \bar{\lambda}_2 \sin \theta ), &  \bar{\lambda}'_2 & = -\bar{\lambda}_1 \sin \theta + \bar{\lambda}_2 \cos \theta , \notag \\
 \eta'_1 & = e^{-i \phi}(\eta_1 \cos \theta + \eta_2 \sin \theta ), &  \eta'_2 & = -\eta_1 \sin \theta + \eta_2 \cos \theta .
 \end{align}
The integration over $\phi$, combined with the homogeneous transformation of $\mathcal{P}$ ensures that the final result transform homogeneously with respect to phase shifts (\ref{eq:phaseshift}) of $\Lambda_1$ of $\Lambda_2$, i.e. that the action of $\gen{D}_{2 \rightarrow 2}$ yields a sum over zero central charge monomials (basis states). When $\gen{D}_{2 \rightarrow 2}$ or analogous expression appear below, we will suppress the phase integration and the phase factors in the change of variables, with the understanding that we only consider terms  that transform homogeneously under the phase shift. 
\section{$\gen{D}_{L \rightarrow 2}$ from scattering amplitudes \label{sec:theproposal}}

The commutation relation between $\dot{\gen{S}}$ and $\gen{D}$ (\ref{eq:sdotcommutatorbybcharge}) closely resembles the annihilation of $\mathcal{N}=4$ SYM tree-level scattering amplitudes by the deformed classical $\bar{\gen{S}}$ superconformal symmetry generator introduced in the work of \cite{Bargheer:2009qu}. Like the $2 \rightarrow 1$ correction here, there the classical $\bar{\gen{S}}$ generator receives a correction that removes one leg of a scattering amplitude generating functional and replaces it with two.  Moreover,   after the translation to spinor helicity coordinates  $\dot{\gen{S}}_{2 \rightarrow 1}$ (\ref{eq:sandsdotinspinorhelicity})  takes a form very close to the expression for $\bar{\gen{S}}_+$ of (3.11)-(3.12) of \cite{Bargheer:2009qu}\footnote{Beside different choice of variable names, the main difference stems from the generators there acting on a generating functional. That form also includes the nonplanar generalization.}. Together with the uniqueness of the solution for $\gen{D}_{L \rightarrow 2}$ and the natural map between the oscillators and the spinor helicity superspace coordinates, this suggests that there is a simple relation  between tree-level scattering amplitudes and the $\gen{D}_{L \rightarrow 2}$. In this section we first write such a relation, and then check it for $L=2$ and $L=3$.

\subsection{The amplitude expression}
We propose that the leading planar $\gen{D}_{L \rightarrow 2}$ is the color-ordered scattering amplitude $A_{L+2}(\Lambda_1 \ldots \Lambda_{L+2})$. Here $A_{L+2}$ refers to the sum over all non-vanishing $N^k \text{MHV}$ $L+2$ particle amplitudes. More precisely,  $\gen{D}_{L \rightarrow 2}$  is the amplitude operator $\mathbf{A}$ that has matrix elements\footnote{Note the reverse ordering of the $\Lambda$ on the left side of the equation.} 
\[
\langle \Lambda_1 \Lambda_2 | \mathbf{A} |\Lambda_{L+2} \Lambda_{L+1} \ldots  \Lambda_3  \rangle = 2 (2 \pi)^{1-2L}\, A_{L+2}(\Lambda^-_1, \Lambda^-_2, \Lambda_3, \Lambda_4, \ldots \Lambda_{L+2}).
\]
Here the minus superscripts for $\Lambda_1$ and $\Lambda_2$ denote negative energy representations, which means that the superconformal generators $\gen{P}$, $\gen{K}$, $\gen{Q}$ and $\gen{S}$ act with opposite sign. The remaining $L$ $\Lambda$ are positive energy. We define the $A_L$ to include no powers of $g$, $2 \pi$ or $i$, absorbing all such necessary prefactors into the $2 (2 \pi)^{1-2 L}$ coefficient. With this normalization we have
\[
A_4(\Lambda_1, \Lambda_2, \Lambda_3, \Lambda_4) = \frac{\delta^4(P) \delta^8(Q)}{\inner{12}\inner{23}\inner{34}\inner{41}}  , \quad   A_5^{\text{MHV}}(\Lambda_i) = \frac{\delta^4(P) \delta^8(Q)}{\inner{12}\inner{23}\inner{34}\inner{45}\inner{51}}.
\]
We then obtain
\begin{eqnarray}
\lefteqn{\langle \Lambda_1 \Lambda_2 | \gen{D}_{L \rightarrow 2} \state{\mathcal{P}_L}  = \langle \Lambda_1 \Lambda_2 | \mathbf{A}\state{\mathcal{P}_L}  =}  \\
& & \! \! 2 (2 \pi)^{1-2L}\, \int \mathrm{d}\Lambda_3 \, \mathrm{d} \Lambda_4 \ldots \mathrm{d}\Lambda_{L+2}  \, A_{L+2}(\Lambda^-_1,\Lambda^-_2, \Lambda_3, \Lambda_4,\ldots \Lambda_{L+2})\, \mathcal{P}_L(\Lambda_{L+2},\Lambda_{L+1}, \ldots \Lambda_3). \notag \label{eq:proposal}
 \end{eqnarray}
To reach the second line we inserted the identity operator (\ref{eq:lambdabasisidentity}). As a first consistency check of this proposal, observe that the helicity ($B$-charge) expansion of $\gen{D}_{L \rightarrow 2}$ of (\ref{eq:bchargeexpansion}) matches the set of nonvanishing $N^k\text{MHV}$ $(L+2)$-particle amplitudes.

Regularization is necessary for $L=2$. As we will see, the singularities occur when the initial and final pairs of $\Lambda$ are equal.   Regularization then requires  adding a (divergent) term proportional to the identity, corresponding to the regularization of the Euler beta function of (\ref{eq:D2to2}). We discuss this further below, when we work out the $L=2$ example.

We will abbreviate the $\Lambda_i$  arguments simply with $i$. For instance,
\begin{eqnarray}
\lefteqn{\langle \Lambda_1, \Lambda_2 | \gen{D}_{L \rightarrow 2} \state{\mathcal{P}_L}  =} \notag \\
& & 2 (2 \pi)^{1-2L}\, \int \mathrm{d}\Lambda_3 \, \mathrm{d} \Lambda_4 \ldots \mathrm{d}\Lambda_{L+2}  \, A_{L+2}(1^-,2^-,3, \ldots L+2)\, \mathcal{P}_L((L+2),(L+1),\ldots 3).
\nln
\end{eqnarray}
%

\subsection{Evaluating the amplitude expression for $\gen{D}_{2\rightarrow 2}$}
As the first example, we check the relation (\ref{eq:proposal}) for $\gen{D}_{2\rightarrow2}$ \footnote{I thank Niklas Beisert for sharing with me his observation that the 4-particle MHV amplitude gives the one-loop dilatation generator in the form given in \cite{Zwiebel:2007th}.}:
\<
\langle \Lambda_1 \Lambda_2 | \gen{D}_{2\rightarrow2}\state{\mathcal{P}_2}  \eq 2 (2 \pi)^{-3} \int \mathrm{d}\Lambda_3\, \mathrm{d}\Lambda_4 \, A_4(1,2,3,4)  \mathcal{P}_2(4,3)
\nln
\eq 2 (2 \pi)^{-3}  \int \mathrm{d}\Lambda_3\, \mathrm{d}\Lambda_4 \, \frac{\delta^4(P) \delta^8(Q)}{\inner{12}\inner{23}\inner{34}\inner{41}}  \mathcal{P}_2(4,3).
\>
To simplify this integration and the integration for $L=3$ below, we will use matrices  $V_{i,j}(\theta)$ for $i < j$, which represent rotations of magnitude $\theta$ in the $i-j$ plane. So $V_{i,j}$ has components
\[ 
(V_{i,j}(\theta))_{k,l} =   \delta_{k,l}(1-\delta_{k,i})(1-\delta_{l,j})   + \cos{\theta}\,( \delta_{k,i}\delta_{l,i} +  \delta_{k,j}\delta_{l,j}) + \sin{\theta}\,( \delta_{k,j}\delta_{l,i} -  \delta_{k,i}\delta_{l,j}). \label{eq:defineVrotation}
\]
Because of the momentum-conservation delta function in $A_4$, it is convenient to change variables as
\<
\begin{pmatrix} \lambda^1_4 \\ \lambda^1_3 \end{pmatrix} \eq r_1 e^{i \sigma_1} U \begin{pmatrix} \lambda^1_1 \\ \lambda^1_2 \end{pmatrix}, \nln
\begin{pmatrix} \lambda^2_4 \\ \lambda^2_3 \end{pmatrix} \eq r_2  \, U \, V_{1,2}(\sigma_2) \begin{pmatrix} \lambda^2_1 \\ \lambda^2_2 \end{pmatrix}, \label{eq:d2changeofvariables}
\>
where $U$ is a unitary matrix, which we parameterize as 
\[
U = \mathrm{diag}(e^{i \phi_2},e^{i \phi_3}) \, V_{1,2}(\theta) \, \mathrm{diag}(1,e^{i \phi_1}).
\]
The $r_i$ range from $0$ to $\infty $,  $\theta$ and $\sigma_2$ from 0 to $\pi/2$, and  $\sigma_1$ and the $\phi_i$  from 0 to $2 \pi$. Since we work in  Lorentzian signature the $\bar{\lambda}$ follow from complex conjugation. Including the minus signs for the momenta of the negative energy representations,
the momentum-conservation delta function localizes at $r_i=1$ and $\sigma_i=0$ :   
\<
\delta^4(P) \eq \delta^4(\sum_{i=1}^4 P_i) =\prod_{\alpha=1}^2\prod_{\dot \beta=1}^2 \delta(-\lambda^\alpha_1 \bar{\lambda}^{\dot{\beta}}_1 -\lambda^\alpha_2 \bar{\lambda}^{\dot{\beta}}_2 +\lambda^\alpha_3 \bar{\lambda}^{\dot{\beta}}_3 + \lambda^\alpha_4 \bar{\lambda}^{\dot{\beta}}_4)
\\
 \eq \frac{i\, \delta(r_1-1)\,\delta(r_2-1)\,\delta(\sigma_1)\,\delta(\sigma_2)}{4 (\sum_{i=1}^2 \lambda_i^1 \bar{\lambda}_i^{\dot{1}} )(\sum_{i=1}^2 \lambda_i^2 \bar{\lambda}_i^{\dot{2}})\big(\inner{12}(\bar{\lambda}_1^{\dot{1}}\bar{\lambda}_1^{\dot{2}}+ \bar{\lambda}_2^{\dot{1}} \bar{\lambda}_2^{\dot{2}}) + (\lambda_1^1\lambda_1^2+\lambda_2^1\lambda_2^2\big)[12])}. \notag
\>
The denominator cancels with  the Jacobians for the change of integration variables, 
\[
\mathrm{d} \Lambda_3 \, \mathrm{d} \Lambda_4 \delta^4(P) \rightarrow  \mathrm{d}^4 \eta_3 \,  \mathrm{d}^4 \eta_4 \, \mathrm{d} \phi_1 \, \mathrm{d} \phi_2 \, \mathrm{d} \phi_3 \,  \mathrm{d} \theta \, (-2 i) \cos \theta \sin \theta,
\]
where now (\ref{eq:d2changeofvariables}) should be used with $r_i=1$, $\sigma_i=0$. 
Evaluating the denominator factor of $A_4$ using this localized change of variables, we find
\begin{eqnarray}
\lefteqn{2 (2 \pi)^{-3} \int \mathrm{d}\Lambda_3 \, \mathrm{d}\Lambda_4 \frac{\delta^4(P)\delta^8(Q)}{\inner{12}\inner{23}\inner{34}\inner{41}}  \mathcal{P}_2(4,3)  =}  \\
& & -4 i (2 \pi)^{-3} \int \mathrm{d}^4 \eta_3 \, \mathrm{d}^4 \eta_4 \, \mathrm{d}\theta \,  \mathrm{d}\phi_1  \, \mathrm{d}\phi_2 \, \mathrm{d}\phi_3 \, e^{-2 i (\phi_1+\phi_2+\phi_3)} \cot \theta  \frac{\delta^8(Q)}{(\inner{12})^4}  \mathcal{P}_2(1',2'), \notag
\end{eqnarray}
where $\mathcal{P}_2$  is evaluated for  $\lambda'=U \lambda$:
\[
\lambda'^\alpha _1= e^{i \phi_2} \Big(\lambda_1^\alpha \cos \theta - e^{i \phi_1} \lambda_2^\alpha \sin \theta \Big)  \quad 
\lambda'^\alpha_2 =e^{i \phi_3} \Big( \lambda_1^\alpha \sin \theta + e^{i \phi_1}\lambda_2^\alpha \cos \theta \Big).
\]
Using the expansion of the fermionic delta function applicable for $\Lambda_1$ and $\Lambda_2$ being negative energy representations\footnote{We use the unit step function $\theta(2-i)$, which is 1 for $i=1,2$ and 0 otherwise.}, 
\[
\delta^8(Q) =  \prod_{A=1}^4 \sum_{j=2}^4 \sum_{i=1}^{j-1}\inner{ij} \eta_i^A\eta_j^A \mathrm{sign}(i,j), \quad \mathrm{sign}(i,j) = (-1)^{\theta(2-i)+\theta(2-j)},
\]
we find that the Grassmann integration of the fermionic delta function divided by $\inner{12}^4$ reduces to the condition that the $\eta$ rotate in parallel to the $\bar{\lambda}$. It is clear then that the $\phi_i$ integrations are simply projections on to a $\mathcal{P}_2$ that transforms homogeneously.
Combining the above results and keeping track of powers of $2 \pi i$ from the $\phi$ integrations, we have
\<
\langle \Lambda_1 \Lambda_2 | \gen{D}_{2\rightarrow2}\state{\mathcal{P}_2}  \eq  2 (2 \pi)^{-3} \int \mathrm{d}\Lambda_3 \, \mathrm{d}\Lambda_4 \, A_4(1^-,2^-,3,4) \, \mathcal{P}_2(4,3)  
\nln
 \eq -4 \int_0^{\pi/2} \mathrm{d}\theta  \, \cot \theta \,  \mathcal{P}_2(1',2'),
\>
where now all $\lambda,$ $\bar{\lambda}$ and $\eta$ rotate in unison as
\[
\begin{array}{l@{=}l@{\hspace{1cm}} l@{=}l} \label{eq:definelambdaprime}
\lambda'^\alpha_1 & \lambda^\alpha_1 \cos \theta -  \lambda^\alpha_2 \sin \theta, &
\lambda'^\alpha_2 &  \lambda^\alpha_1 \sin \theta + \lambda^\alpha_2 \cos \theta, \\
\bar{\lambda}'^{\dot{\alpha}}_1& \bar{\lambda}^{\dot{\alpha}}_1 \cos \theta -  \bar{\lambda}^{\dot{\alpha}}_2 \sin \theta, &
\bar{\lambda}'^{\dot{\alpha}}_2 &  \bar{\lambda}^{\dot{\alpha}}_1 \sin \theta + \bar{\lambda}^{\dot{\alpha}}_2 \cos \theta,\\
\eta'^A_1 & \eta^A_1 \cos \theta -  \eta^A_2 \sin \theta, &
\eta'_{2,A} &  \eta^A_1 \sin \theta + \eta^A_2 \cos \theta,
\end{array} 
\]%
and we only keep terms that transform homogeneously when  $\Lambda_1$ or $\Lambda_2$ undergo the phase shift. 

As noted above, this integral has a divergence due to the $\Lambda'$ becoming collinear with the  $\Lambda$ at $\theta=0$.  As we will see below, evaluating $\gen{D}_{3 \rightarrow 2}$ from our scattering amplitude proposal requires no regularization.  Then  we find (numerically) that requiring\footnote{Recall that this commutator only vanishes up to generalized gauge transformations $\dot{\gen{s}}$ which vanish on cyclic states. Once  $\dot{\gen{S}}_{1 \rightarrow 1}$, $\dot{\gen{S}}_{2 \rightarrow 1}$, $\gen{D}_{3 \rightarrow 2}$ and off-diagonal $\gen{D}_{2 \rightarrow 2}$  are fixed, there is a unique solution of (\ref{eq:dscommutator}) for both the (diagonal) regularization of  $\gen{D}_{2 \rightarrow 2}$ and  $\dot{\gen{s}}$.}
\[
\comm{\dot{\gen{S}}_{2 \rightarrow 1}}{\gen{D}_{2 \rightarrow 2}} + \comm{\dot{\gen{S}}_{1 \rightarrow 1}}{\gen{D}_{3 \rightarrow 2}} =0 \label{eq:dscommutator}
\]
fixes the regularization of $\gen{D}_{2 \rightarrow 2}$ uniquely. It would be nice to find a (more) physical argument for the precise regularization in addition to this argument from superconformal symmetry. Including the regularization, we finally have 
 \[
\langle \Lambda_1 \Lambda_2 |\gen{D}_{2 \rightarrow 2}\state{\mathcal{P}_2} = 4  \int \mathrm{d} \theta \,  \cot \theta \Big(\mathcal{P}_2(\Lambda_1,\Lambda_2) - \mathcal{P}_2(\Lambda'_1,\Lambda'_2) \Big),
 \]
 where the $\Lambda'$ are defined in (\ref{eq:definelambdaprime}). This agrees with (\ref{eq:d2insuperspace}), verifying our proposed relation with scattering amplitudes for $L=2$.
\subsection{Evaluating the amplitude expression for $\gen{D}_{3\rightarrow 2}$}
In this section we evaluate $\gen{D}_{3\rightarrow 2}$ using similar steps to those used above. This will provide additional confirmation of the relation  (\ref{eq:proposal}) to scattering amplitudes. That relation gives $\gen{D}_{3\rightarrow 2}$  in terms of  five-particle scattering. For five particles, the nonvanishing $N^k$MHV amplitudes have $k=0,1$. Since $k=1$ is the $\overline{\text{MHV}}$ amplitude, it is sufficient to compute the contribution from the 5-particle MHV amplitude. Then the  $k=1$, $\overline{\text{MHV}}$ contribution  follows from interchanging $\lambda$ and $\bar{\lambda}$, and $\bar{\eta}$ and $\eta$ (or $\mathbf{a}$ and $\mathbf{b}$, and $\mathbf{c}$ and $\mathbf{d}$). 

 For the calculation of the 5-particle MHV contribution it is easier to work with $\bar{\eta}$ (\ref{eq:introduceetabar}) than $\eta$. The reason is that while the number of $\eta$ are changed by this interaction, the number of $\bar{\eta}$ is conserved.  So using $\bar{\Lambda}$,  the amplitude expression becomes
\<
\langle \bar{\Lambda}_1 \bar \Lambda_2 |\gen{D}^{[0]}_{3\rightarrow2}\state{\mathcal{P}_3} \eq 2 (2 \pi)^{-5} \int \mathrm{d}\bar\Lambda_3\, \mathrm{d}\bar\Lambda_4 \, \mathrm{d}\bar\Lambda_5 \, A_5^{\text{MHV}}(\bar1^-,\bar2^-,\bar3,\bar4,\bar5) \, \mathcal{P}_3(\bar5,\bar4,\bar3)
\\
\eq  2 (2 \pi)^{-5} \int \mathrm{d}\bar\Lambda_3\, \mathrm{d}\bar\Lambda_4 \,  \mathrm{d}\bar\Lambda_5 \, \frac{\delta^4(P) \delta^8(Q)}{\inner{12}\inner{23}\inner{34}\inner{45}\inner{51}}   \mathcal{P}_3(\bar5,\bar4,\bar3). \notag
\>
This time we change variables as
\<
\begin{pmatrix} \lambda_5^1 \\ \lambda_4^1 \\ \lambda_3^1 \end{pmatrix} \eq r_1 e^{i \sigma_1} \, U \begin{pmatrix} \lambda_1^1 \\ \lambda_2^1 \end{pmatrix}, \nln
\begin{pmatrix} \lambda_5^2 \\ \lambda_4^2 \\ \lambda_3^2 \end{pmatrix} \eq r_2  \, U \, V_{1,2}(\sigma_2) \begin{pmatrix} \lambda_1^2 \\ \lambda_2^2 \end{pmatrix}.
\>
$U$ is parameterized now as
\[
U = \mathrm{diag}(e^{i \phi_2},e^{i \phi_3},e^{i \phi_4}) V_{2,3}(\theta_3)V_{1,2}(\theta_2) \, \mathrm{diag}\{1,1, e^{i \rho}) \, U_0 \, \mathrm{diag}\{1,e^{i \phi_1}\},
\]
the $V$ are defined by (\ref{eq:defineVrotation}) and the 3 by 2 matrix $U_0$ is
\[
U_0 = \begin{pmatrix} 1 & 0  \\ 0 & \cos \theta_1 \\ 0 & \sin \theta_1 \end{pmatrix}.
\]
The $r_i$ range from $0$ to $\infty $,  $\sigma_2$ and the $\theta_i$ from 0 to $\pi/2$, and  $\rho$, $\sigma_1$ and the $\phi_i$ from 0 to $2 \pi$. Again the momentum delta function imposes $r_i=1$ and $\sigma_i=0$, and the bosonic integration localizes on the other eight (bosonic) variables. As for $\gen{D}_{2\rightarrow 2}$, we compute the Jacobian from the delta function and for the change of variables, and we evaluate the product of spinor products in the denominator in the new variables.
Now the Grassmann integration reduces to the condition that the $\bar{\eta}$ rotate in the same way as the $\lambda$. Again  the $\phi_i$ integrations are simply projections enforcing homogeneous transformations under the phase shift. 
 Finally, we obtain 
\begin{eqnarray}
\langle \bar{\Lambda}_1 \bar \Lambda_2 |\gen{D}^{[0]}_{3\rightarrow2}\state{\mathcal{P}_3} \eq  2 (2 \pi)^{-5}\int \mathrm{d}\bar\Lambda_3 \, \mathrm{d}\bar\Lambda_4  \, \mathrm{d}\bar\Lambda_5 \, A^{\text{MHV}}_5(\bar1^-,\bar2^-,\bar3,\bar4,\bar5) \mathcal{P}_3(\bar5,\bar4,\bar3) 
\nln
\eq   \frac{16}{2 \pi i}  \int \mathrm{d}\rho\prod_{i=1}^3 \mathrm{d}\theta_i \frac{c_2}{c_1} \frac{e^{-i \rho}}{1 - e^{i\rho} \frac{s_1 c_2 s_3}{c_1 c_3}}  [12]  \mathcal{P}_3(\bar1',\bar2',\bar3'), \label{eq:d3to2}
\end{eqnarray}
where we have used the abbreviations $c_i=\cos \theta_i$ and $s_i=\sin \theta_i$, and now all $\lambda,$ $\bar{\lambda}$ and $\bar{\eta}$ rotate in unison (up to complex conjugation for $\bar{\lambda}$) as 
\[
\begin{array}{lcl@{\hspace{-.05cm}} lcl }
\lambda'^\alpha_1 &=& \lambda^\alpha_1 c_2 -  \lambda^\alpha_2 c_1 s_2 & \bar{\eta}'_{1,A} &=& \bar{\eta}_{1,A} \, c_2 -  \bar{\eta}_{2,A} \,  c_1 s_2 , \\
\lambda'^\alpha_2  & =&  \lambda^\alpha_1  s_2 c_3 + \lambda^\alpha_2  (c_1 c_2 c_3 - e^{i \rho} s_1 s_3) & \bar{\eta}'_{2,A} &=& \bar{\eta}_{1,A} \, s_2 c_3  + \bar{\eta}_{2,A} (c_1 c_2 c_3 - e^{i \rho} s_1 s_3),  \\
 \lambda'^\alpha_3  & = &  \lambda^\alpha_1  s_2 s_3 + \lambda^\alpha_2  (c_1 c_2 s_3 + e^{i \rho} s_1 c_3) &  \bar{\eta}'_{3,A} &=&    \bar{\eta}_{1,A}\,  s_2 s_3 +  \bar{\eta}_{2,A} (c_1 c_2 s_3 + e^{i \rho} s_1 c_3), \\
 \bar{\lambda}'^{\dot{\alpha}}_1 &=& \bar{\lambda}^{\dot{\alpha}}_1 c_2 -  \bar{\lambda}^{\dot{\alpha}}_2 c_1 s_2 & \bar\lambda'^{\dot{\alpha}}_2 & =&  \bar{\lambda}^{\dot{\alpha}}_1 s_2 c_3 + \bar{\lambda}^{\dot{\alpha}}_2 (c_1 c_2 c_3 - e^{-i \rho} s_1 s_3),  \\
 \bar{\lambda}'^{\dot{\alpha}}_3 & = &  \bar{\lambda}^{\dot{\alpha}}_1 s_2 s_3 + \bar{\lambda}^{\dot{\alpha}}_2 (c_1 c_2 s_3 + e^{-i \rho} s_1 c_3). & & &
\end{array} 
\]
As usual, instead of explicitly writing the $\phi_i$ integrations, we only keep terms that transform homogeneously. Expanding $\mathcal{P}_3(\Lambda'_1,\Lambda'_2,\Lambda'_3)$ in powers of $e^{i \rho}$, it is straightforward to evaluate the $\rho$ integral, yielding sums of terms with coefficients that are $\theta_i$-integrals of two types\footnote{Only odd powers of cosines and sines appear due to the projection onto homogeneous terms. We choose to write the odd powers of cosines as a single cosine times powers of sines.}:
\<
I_1(i,j,k) \eq 8 \int_0^{\pi/2} \mathrm{d}\theta_1   \int_0^{\pi/2} \mathrm{d}\theta_2 \int_0^{\pi/2} \mathrm{d}\theta_3 \, \theta(\frac{c_1 c_3}{s_1 s_3} - c_2)(s_1^{2 i +1} c_1)( s_2^{2 j +1} c_2)( s_3^{2 k+1} c_3),
\nln
I_2(i,j,k) \eq 8 \int_0^{\pi/2} \mathrm{d}\theta_1   \int_0^{\pi/2} \mathrm{d}\theta_2 \int_0^{\pi/2} \mathrm{d}\theta_3 \, \theta(c_2-\frac{c_1 c_3}{s_1 s_3})(s_1^{2 i +1} c_1)( s_2^{2 j +1} c_2)( s_3^{2 k+1} c_3).
\nln
\>
$\theta$ is the  Heaviside theta function.  The $I_1$ integrals appear only with $i,j,k \geq 0$, while the $I_2$ appear with $i,j \geq 0$ and $k \geq -1$.  These integrals can be evaluated analytically: 
\<
I_1(i,j,k) \eq \frac{1}{i+1}\frac{1}{j+1}B(i+2,k+1) + \sum_{i_2=i}^{i+k+1} (-1)^{i+i_2+k}  \binom{i+1}{i_2-k} \binom{i_2}{i}\frac{\Delta_S(i_2,j)}{i+1}, \nln
I_2(i,j,k) \eq \frac{1}{i+1}\frac{1}{j+1}\frac{1}{k+1} - I_1(i,j,k) \quad (k\geq0), \\
I_2(i,j,-1) \eq \frac{\Delta_S(i,j)}{j+1}, \nln
\Delta_S(i,i) \eq \zeta(2) - S_2(i+1), \quad  \Delta_S(i,j) = \frac{S_1(i+1)-S_1(j+1)}{i-j} \quad (i \neq j).\notag
\>
$B(i,k)$ is the Euler beta function, $\zeta(2)=\pi^2/6$,  $S_1(n)$ is the $n$th harmonic number, and $S_2(n)$ is the $n$th harmonic number of second order as defined in (\ref{eq:defineharmonics}).   Importantly, we find that this expression (\ref{eq:d3to2}) for $\gen{D}_{3 \rightarrow 2}$ is finite since it is always given as a finite sum of the $I_1$ and $I_2$ {}\footnote{After doing the $\rho$ integration, one should combine terms completely before doing the $\theta_i$ integrations, since generically some terms with negative powers of the $s_i$ cancel inside the integrand.}. While the $I_j$ integrals have terms proportional to $\zeta(2)$, these terms always cancel in (\ref{eq:d3to2}).
 
 With the help of $\texttt{Mathematica}$, it is straightforward to compute the action of this expression for $\gen{D}_{3\rightarrow 2}$ (\ref{eq:d3to2}) on arbitrary three site states (with not too many powers of  $\lambda$ and $\bar\lambda$). We have checked for many random initial and final states with up to $6$ powers of $\bar{\lambda}$ initially that this expression agrees with the solution we derived earlier (\ref{eq:kidentity}-\ref{eq:parityandconjugation}). This numerical confirmation complements the more analytic approach we will pursue in the next section.

\section{Checking superconformal symmetry \label{sec:superconformal}}

In this section we assume that the proposed form of $\gen{D}_{L \rightarrow 2}$ in terms of scattering amplitudes (\ref{eq:proposal}), combined with the regularization for $L=2$,  maps any homogeneous polynomial in  of $\lambda$, $\bar{\lambda}$ and $\bar \eta$ to another such polynomial. At the end of this section, we will discuss evidence in favor of  this assumption.

Under this assumption we will prove\footnote{For $L=3$ we will rely on numerical confirmation.} that the amplitude expression for $\gen{D}_{L \rightarrow 2}$ satisfies the constraints from superconformal symmetry that uniquely fix $\gen{D}_{L \rightarrow 2}$  in terms of $\gen{D}_{L' \rightarrow 2}$ with $L' < L$. Combined with the matching for $L=2$, this then proves this relation between scattering amplitudes and the dilatation generator.  

Recall that the superconformal symmetry constraints reduce to the condition that $\gen{D}_{L \rightarrow 2}$ commute with all $\mathcal{O}(g^0)$ $\alg{psu}(2,2|4)$ symmetry generators. So, we will first show that the superconformal symmetry generators with only $1 \rightarrow 1$ interactions commute with the amplitude definition for $\gen{D}_{L \rightarrow 2}$ (\ref{eq:proposal}). To do this we show that the commutator with the symmetry generators is proportional to the symmetry generator acting on the amplitude, which vanishes. Section \ref{sec:2to1} checks commutators with the symmetry generators  that also have  $2 \rightarrow 1$ interactions. We show that a conformal anomaly contribution for the commutator with the $1 \rightarrow 1$ part of the symmetry generator  is canceled by the commutator with the $2 \rightarrow 1$ part of the generator. 

This section's calculations are  similar to the calculations of \cite{Bargheer:2009qu}, which is not surprising since the deformations of \cite{Bargheer:2009qu} are closely related to our $2 \rightarrow 1$ interactions, and these deformations were constructed to cancel conformal anomaly contributions.

\subsection{Commutators with length-preserving $\alg{psu}(2,2|4)$ generators \label{sec:1to1commutators}}
We first focus on $\alg{psu}(2,2|4)$ generators  beside $\gen{S}$, $\dot{\gen{S}}$ and $\gen{K}$, which have only $1\rightarrow 1$ interactions (\ref{eq:classicalgeneratorinsuperspace}) at leading order (in the coupling constant convention of this work).  We will examine one simple example in full detail. This should suffice to show how all of these constraints are satisfied. We consider the commutator of $\gen{L}_\beta^\alpha (\alpha \neq \beta)$ with $\gen{D}_{2 \rightarrow 2}$:
\<
\langle\Lambda_1\Lambda_2|\comm{\gen{L}_\beta^\alpha}{\gen{D}_{2\rightarrow2}}\state{\mathcal{P}_2} & \propto & \int \mathrm{d}\Lambda_3\, \mathrm{d}\Lambda_4  \sum_{i=1}^2 \lambda^\alpha_i \partial_{i,\beta} A_4(1^-,2^-,3,4) \mathcal{P}_2(4,3)  
\nl
-  \int \mathrm{d}\Lambda_3\, \mathrm{d}\Lambda_4 \, A_4(1^-,2^-,3,4) \sum_{i=3}^4 \lambda^\alpha_i \partial_{i,\beta} \mathcal{P}_2(4,3) 
\nln
\eq  \int \mathrm{d}\Lambda_3\, \mathrm{d}\Lambda_4  \sum_{i=1}^4 \big( \lambda^\alpha_i \partial_{i,\beta} A_4(1^-,2^-,3,4) \big) \mathcal{P}_2(4,3)  
\nln
\eq  \int \mathrm{d}\Lambda_3\, \mathrm{d}\Lambda_4  \, \big( \gen{L}_\beta^\alpha A_4(1^-,2^-,3,4) \big) \mathcal{P}_2(4,3)  
\nln
\eq 0.
\>
For the first term of the commutator $\gen{L}_\beta^\alpha$ acts after $\gen{D}_{2 \rightarrow 2}$, and therefore acts on $\Lambda_1^-$ and $\Lambda_2^-$. For the second term of the commutator (with a minus sign), $\gen{L}$ acts first on $\mathcal{P}$, which is equivalent to acting on $\Lambda_3$ and $\Lambda_4$.
The third line then follows from integration by parts (and combining the two terms on the right side of the first two lines). The integrand in the third line then includes a factor of $\gen{L}_\beta^\alpha$ acting on the amplitude, which is zero. 

The remaining commutators with $\gen{D}_{2 \rightarrow 2}$ work similarly. Integration by parts and the negative energy representations of  $\Lambda^-_1$ and $\Lambda^-_2$ imply that the commutator with a generator $\gen{J}$  is proportional to an integral of  $\gen{J}$ acting on a scattering amplitude.  Moreover, this check would work the same way for $\gen{D}_{L \rightarrow 2}$ since both in the commutator and in the action on the amplitude we sum over all $L$ positive energy representations.
In other words, for such $\alg{psu}(2,2|4)$ generators $\gen{J}$ with only $1 \rightarrow 1$ interactions,   we obtain
\< \label{eq:general1sitecommutator}
\lefteqn{\langle\Lambda_1\Lambda_2|\comm{\gen{J}}{\gen{D}_{L\rightarrow2}}\state{\mathcal{P}_2} \propto}
\\
& &    \int \mathrm{d}\Lambda_3\, \mathrm{d}\Lambda_4  \ldots \mathrm{d}\Lambda_{L+2} \big( \gen{J} A_{L+2}(1^-,2^-, 3, 4, \ldots (L+2)) \big) \mathcal{P}_L((L+2),(L+1), \ldots 1) , \notag
\>
which is zero. Similarly, the commutator with $\gen{C}$ vanishes as required, and the commutator with $\gen{B}$ confirms the mapping between $N^k\text{MHV}$ amplitudes and $\gen{D}_{L \rightarrow 2}^{[k]}$ interactions noted earlier. 

\subsection{Commutators with length-changing $\alg{psu}(2,2|4)$ generators \label{sec:2to1}}

In this section we consider commutators between $\gen{D}_{L \rightarrow 2}$ with the remaining generators: $\gen{S}$, $\dot{\gen{S}}$ and $\gen{K}$. Since $\gen{S}$ and $\dot{\gen{S}}$ anticommute to $\gen{K}$ it suffices to show that the commutators with $\gen{S}$ and $\dot{\gen{S}}$ vanish.  As in the previous section, we will consider one example in full detail, and then explain why these commutators vanish in general.

We will show that the scattering amplitude-dilatation generator relation is consistent with (recall (\ref{eq:sdotcommutatorbybcharge}))
\[   \label{eq:sdotdcommutator}
 \comm{(\dot{\gen{S}}_{\dot{\alpha}}^B)_{1 \rightarrow 1}}{\gen{D}^{[0]}_{4\rightarrow 2}} + \comm{(\dot{\gen{S}}_{\dot{\alpha}}^B)_{2 \rightarrow 1}}{\gen{D}^{[0]}_{3\rightarrow 2}}
     = 0.
\]
As in (\ref{eq:general1sitecommutator}), the first term is equivalent to an integral which includes a factor of a classical $1 \rightarrow 1$ generator acting on an amplitude. However, recalling that $\dot{\gen{S}}_{\dot{\alpha}}^B=\partial_{\dot{\alpha}} \eta^B$ (\ref{eq:classicalgeneratorinsuperspace}), the first term of (\ref{eq:sdotdcommutator})  is not zero due to the conformal anomaly \cite{Cachazo:2004by},
\[ \label{eq:conformalanomaly}
\partial_{i,\dot{\alpha}} \frac{1}{\inner{ij}} =  \pi \varepsilon_{\dot{\alpha}\dot{\beta}}\bar{\lambda}_j^{\dot{\beta}}\, \delta^2(\inner{ij}).
\]
Using (\ref{eq:proposal}) to write matrix elements of $\gen{D}^{[0]}_{4\rightarrow 2}$ as
\[
\langle \Lambda_1 \Lambda_2 |\gen{D}^{[0]}_{4\rightarrow 2} \state{\mathcal{P}_4}= 2 (2 \pi)^{-7} \int \mathrm{d}\Lambda_3 \, \mathrm{d}\Lambda_4 \, \mathrm{d}\Lambda_5 \, \mathrm{d}\Lambda_6 \, A_6^{\text{MHV}}(1^-,2^-,3,4,5,6) \mathcal{P}_4(6,5,4,3),
\]
we can write the conformal anomaly as a sum over six terms, each one proportional to one of the $\delta^2(\inner{i(i+1)})$ for $i=1,2,\ldots 6$.  The term proportional to $\delta^2(\inner{12})$ does not contribute since we are free to assume that $\lambda_1$ and $\lambda_2$ are not collinear. So we are left with five terms due to the conformal anomaly. 

On the other hand, the second term of the original commutator (\ref{eq:sdotdcommutator}) has the following five contributions from (\ref{eq:sdeltadcommutator}),
\begin{gather}
\label{eq:sdot2to1d3to2commutator}  (\dot{\gen{S}}_{\dot{\alpha}}^B)_{2 \rightarrow 1}(2,3) \gen{D}^{[0]}_{3\rightarrow 2}(1,2,3) +  (\dot{\gen{S}}_{\dot{\alpha}}^B)_{2 \rightarrow 1}(1,2) \gen{D}^{[0]}_{3\rightarrow 2}(2,3,4)  
\\
  -  \gen{D}^{[0]}_{3\rightarrow 2}(1,2,3)(\dot{\gen{S}}_{\dot{\alpha}}^B)_{2 \rightarrow 1}(1,2)-  \gen{D}^{[0]}_{3\rightarrow 2}(1,2,3)(\dot{\gen{S}}_{\dot{\alpha}}^B)_{2 \rightarrow 1}(2,3) -   \gen{D}^{[0]}_{3\rightarrow 2}(1,2,3)(\dot{\gen{S}}_{\dot{\alpha}}^B)_{2 \rightarrow 1}(3,4). \notag
 \end{gather}
There is a one-to-one cancellation between terms. In order, the two terms of the first line of (\ref{eq:sdot2to1d3to2commutator}) cancel the $\delta^2(\inner{23})$ and $\delta^2(\inner{61})$ conformal anomaly terms. Similarly, in order, the three terms of the second line of  (\ref{eq:sdot2to1d3to2commutator}) cancel the $\delta^2(\inner{56})$,  $\delta^2(\inner{45})$, and $\delta^2(\inner{34})$ conformal anomaly terms. We now work through one cancellation from each line, and the other three can be done similarly.

For the first term of (\ref{eq:sdot2to1d3to2commutator}), combining the relation to scattering amplitudes (\ref{eq:proposal}) for $L=3$ and the spinor-helicity superspace expression for $\dot{\gen{S}}$ of (\ref{eq:sandsdotinspinorhelicity}) gives
\< \lefteqn{ \langle\Lambda_1 \Lambda_2 | (\dot{\gen{S}}_{\dot{\alpha}}^B)_{2 \rightarrow 1}(2,3) \gen{D}^{[0]}_{3\rightarrow 2}(1,2,3)\state{\mathcal{P}_4} =} \label{eq:firsttermofsdot2to1d3to2commutator}
\\
& & 
2 (2 \pi)^{-5} \, \varepsilon_{\dot{\alpha}\dot{\beta}}  \bar{\lambda}^{\dot{\beta}}_2  \int \mathrm{d}^4 \eta'  \eta'^B \int_0^{\pi/2} \mathrm{d} \theta  \int \mathrm{d} \Lambda_4 \, \mathrm{d} \Lambda_5 \,  \mathrm{d} \Lambda_6 \, A_5^{\text{MHV}}(1^-,2'^-,4,5,6)\mathcal{P}_4(6,5,4,3'). \notag
\>
The new variables are defined as
\[
\begin{array}{l@{=}l@{\hspace{1cm}} l@{=}l}
\lambda'^\alpha_2& \lambda^\alpha_2 \cos \theta, &\lambda'^\alpha_3&\lambda^\alpha_2 \sin \theta,  \\
\bar{\lambda}'^{\dot{\alpha}}_2& \dot{\lambda}^{\dot{\alpha}}_2 \cos \theta, &\bar{\lambda}'^{\dot{\alpha}}_3&\bar{\lambda}^{\dot{\alpha}}_2 \sin \theta, \\
\eta'^A_2&\eta^A_2 \cos \theta - \eta'^A \sin\theta, & \eta'^A_3 & \eta_2^A \sin \theta + \eta'^A \cos \theta.
\end{array}
\]
The corresponding conformal anomaly term proportional to $\delta^2(\inner{23})$ is given by
\[
2(2 \pi)^{-7} \,  \pi \varepsilon_{\dot{\alpha}\dot{\beta}} \int \mathrm{d} \Lambda_3 \, \mathrm{d} \Lambda_4 \,  \mathrm{d} \Lambda_5 \, \mathrm{d} \Lambda_6 \,  (\bar{\lambda}'^{\dot{\beta}}_3\,\eta^B_2-\bar{\lambda}^{\dot{\beta}}_2\eta'^B_3) \mathcal{P}_4(6,5,4,3) \frac{ \delta^2(\inner{23})\delta^4(P)\delta^8(Q)}{\inner{12}\inner{34}\inner{45}\inner{56}\inner{61}}
\]
We change variables as \cite{Bargheer:2009qu}
\[
\lambda^\alpha_3 = e^{i \phi} \lambda^\alpha_2 \sin \theta +  \lambda'^\alpha z, \quad \bar{\lambda}^{\dot{\alpha}}_3 = e^{-i \phi} \bar{\lambda}^{\dot{\alpha}}_2 \sin \theta +  \bar \lambda'^{\dot{\alpha}} \bar{z}, \quad \eta^A_3 = e^{-i \phi}  \eta^A_2 \sin \theta + \eta'^A \cos \theta,
\]
where $\lambda'$ is a constant  spinor. Here we can use $\theta$ running from $0$ to $\pi/2$ because the combined momentum of collinear $2$ and $3$ must still give a negative energy contribution (for there to be nonvanishing support for the delta function of total momentum). Applying this  change of variables,  integrating out the $\delta^2(\inner{23})$  factor, and simplifying\footnote{The $\phi$ integral translates simply to the central charge condition for $3'$.} yields precisely minus of (\ref{eq:firsttermofsdot2to1d3to2commutator}) as claimed.

We repeat the above steps for one other term, the first term on the second line of (\ref{eq:sdot2to1d3to2commutator}). We have
\< \lefteqn{ \langle\Lambda_1 \Lambda_2 |  (-)\gen{D}^{[0]}_{3\rightarrow 2}(1,2,3)(\dot{\gen{S}}_{\dot{\alpha}}^B)_{2 \rightarrow 1}(1,2)\state{\mathcal{P}_4} =} \label{eq:firsttermonsecondlineofsdot2to1d3to2commutator}
\\ 
& & 
-2 (2 \pi)^{-7} \,   \int  \mathrm{d} \Lambda_3 \,  \mathrm{d} \Lambda_4 \,  \mathrm{d} \Lambda_{56} \, \varepsilon_{\dot{\alpha}\dot{\beta}} \bar{\lambda}^{\dot{\beta}}_{56}\int \mathrm{d}^4 \eta' \int_0^{\pi/2} \mathrm{d} \theta \,  \eta'^B \mathcal{P}_4(1',2',4,3) A_5^{\text{MHV}}(1^-,2^-,3,4,56). \notag
\>
This time the new variables are
\[
\begin{array}{l@{=}l@{\hspace{1cm}} l@{=}l}
\lambda'_1& \lambda_{56} \cos \theta, &\lambda'_2&\lambda_{56} \sin \theta,  \\
\bar{\lambda}'_1& \bar{\lambda}_{56} \cos \theta, &\bar{\lambda}'_2&\bar{\lambda}_{56} \sin \theta, \\
\eta'_1&\eta_{56} \cos \theta - \eta' \sin\theta, & \eta'_2 & \eta_{56} \sin \theta + \eta' \cos \theta.
\end{array}
\]
The corresponding conformal anomaly term is proportional to $\delta^2(\inner{56})$,
\[
2 ( 2 \pi)^{-7} \pi \varepsilon_{\dot{\alpha}\dot{\beta}} \int \mathrm{d} \Lambda_3 \, \mathrm{d} \Lambda_4 \, \mathrm{d} \Lambda_5 \, \mathrm{d} \Lambda_6 \, (\bar{\lambda}^{\dot{\beta}}_6\,\eta^B_5-\bar{\lambda}^{\dot{\beta}}_5\eta_6^B)\mathcal{P}_4(6,5,4,3) \frac{ \delta^2(\inner{56})\delta^4(P)\delta^8(Q)}{\inner{12}\inner{23}\inner{34}\inner{56}\inner{61}}.
\]
As in the previous case, we can then match minus (\ref{eq:firsttermonsecondlineofsdot2to1d3to2commutator})  by  changing variables, integrating out the $\delta^2$ factor and simplifying. The change of variables is
\[
\begin{array}{l@{=}l@{\hspace{1cm}} l@{=}l}
\lambda'_1& e^{i \phi} \lambda_{56} \cos \theta, &\lambda'_2& \lambda_{56} \sin \theta  +  \lambda' z, \\
\bar{\lambda}'_1& e^{-i \phi}\bar{\lambda}_{56} \cos \theta, &\bar{\lambda}'_2&\bar{\lambda}_{56} \sin \theta  +  \bar\lambda' \bar{z}, \\
\eta'_1& e^{-i \phi} (\eta_{56} \cos \theta - \eta' \sin\theta), & \eta'_2 & \eta_{56} \sin \theta + \eta' \cos \theta.
\end{array}
\]

Now that we have seen how the $\dot{\gen{S}}$ commutator works for this example, it is straightforward to generalize to all such commutators involving $\gen{D}_{L \rightarrow 2}$ (for $L \geq 3$ only), including all helicity contributions. As reviewed in \cite{Bargheer:2009qu}, when particles (adjacent in the planar limit) become collinear, their singular behavior is governed by universal splitting functions \cite{Berends:1989aa,Mangano:1990by}. Therefore, this  term by term cancellation between the $\dot{\gen{S}}_{2 \rightarrow 1}$ commutator terms and the conformal anomaly terms works for general $L$ and general helicity. More specifically the terms proportional to 
\[
\delta^2\big(\big\langle(i+2)(i+3)\big\rangle\big), \quad i = 1 \ldots L-1
\]
cancel the terms of the form
\[
(-)\gen{D}_{(L-1)\rightarrow 2}\big(1,2,\ldots(L-1)\big)(\dot{\gen{S}}_{\dot{\alpha}}^B)_{2 \rightarrow 1}\big((L-i),(L+1-i)\big).
\]
Similarly, we have\footnote{Also, the $\delta^2(\inner{12})$ term does not contribute as before since we are free to choose arbitrary $\lambda_1$ and $\lambda_2$.}
\begin{eqnarray}
\delta^2(\inner{23}) \text{ term} & +  & (\dot{\gen{S}}_{\dot{\alpha}}^B)_{2 \rightarrow 1}(2,3) \gen{D}_{(L-1)\rightarrow 2}\big(1,2,\ldots(L-1)\big) =0, \notag \\
\delta^2\big(\big \langle(L+2)1\big \rangle \big) \text{ term} & + & (\dot{\gen{S}}_{\dot{\alpha}}^B)_{2 \rightarrow 1}(1,2) \gen{D}_{(L-1)\rightarrow 2}(2,3,\ldots L) = 0.
\end{eqnarray}
Since $\dot{\gen{S}}$ and $\gen{S}$ are exchanged when we interchange  $\lambda$ and $\bar{\lambda}$ and $\eta$ and $\bar{\eta}$, we also see that the superconformal constraint from $\gen{S}$ is satisfied, as we aimed to show.

However, in this subsection so far we have not discussed the commutator 
\[
 \comm{(\dot{\gen{S}}_{\dot{\alpha}}^B)_{2 \rightarrow 1}}{\gen{D}_{2\rightarrow 2}} + \label{eq:sdotdcommutatorlowestorder}
  \comm{(\dot{\gen{S}}_{\dot{\alpha}}^B)_{1 \rightarrow 1}}{\gen{D}_{3\rightarrow 2}}   = 0.
\]
Because we worked out and confirmed numerically the exact analytic expression for $\gen{D}_{3\rightarrow 2}$ (\ref{eq:d3to2}) (and its conjugate), this is less important for checking our proposal. Nonetheless, it would be best to have a complete analytic proof that the constraints from superconformal symmetry are satisfied. However, there are four nongeneric contributions to (\ref{eq:sdotdcommutatorlowestorder}). The first two can be seen in (\ref{eq:sdeltadcommutator}). As noted earlier, this commutator is zero only on cyclic states, and commutes to the gauge transformation $\dot{\gen{s}}$ locally. Second, since $\dot{\gen{S}}_{2 \rightarrow 1}$ only inserts one $\bar{\lambda}$ and $\gen{D}_{2 \rightarrow 2}$ does not insert any, for this commutator only we have contributions where $\dot{\gen{S}}_{2 \rightarrow 1}$ acts on the same sites as $\gen{D}$ (otherwise it acts on one site not acted on by $\gen{D}$). Third, as explained above, the expression for $\gen{D}_{2 \rightarrow 2}$ from $A_4$ using (\ref{eq:proposal}) requires regularization. Fourth and finally, when simplifying the conformal anomaly terms for this commutator, we cannot integrate out the $\delta^2$ first because this also leads to a divergence. The correct way to simplify the conformal anomaly terms is to first evaluate $\gen{D}_{3\rightarrow 2}$ (localizing the delta function of momentum), and then to integrate out the $\delta^2$. It is straightforward to write these four contributions analytically. While simplifying the sum of these four contributions may require lengthy calculation, it should be tractable, and based on the numerical results it is clear that these four special contributions would cancel. It is worth noting, that if we look at the commutator matrix elements (in oscillator notation)
\[
\langle \vec{n}_1\vec{n}_2|\bigg( \comm{(\dot{\gen{S}}_{\dot{\alpha}}^B)_{2 \rightarrow 1}}{\gen{D}_{2\rightarrow 2}} +   \comm{(\dot{\gen{S}}_{\dot{\alpha}}^B)_{1 \rightarrow 1}}{\gen{D}_{3\rightarrow 2}}\bigg) \state{\vec{m}_1 \vec{m}_2\vec{m}_3} \quad \vec{n}_1 \neq \vec{m}_1 \text{ and } \vec{n}_2 \neq \vec{m}_3,
\]
all of these types of contributions vanish\footnote{For the fourth contribution, as long as we restrict our attention to such matrix elements it is ok to integrate out the $\delta^2$ function first since there is no divergence.}. So the missing analytic check is only for such matrix elements where $\vec{n}_1 = \vec{m}_1$ or $\vec{n}_2 = \vec{m}_3$.

We finish with a discussion of our assumption that the proposed mapping from scattering amplitudes to the dilatation generator maps a homogeneous polynomial $\mathcal{P}_L$ of  $\lambda$, $\bar\lambda$, and $\eta$ to another (finite)  polynomial. The proof of superconformal symmetry of this section and the $L=2,3$ examples of the previous section\footnote{We also have worked out the amplitude expression for $\gen{D}^{[0]}_{4 \rightarrow 2}$, confirming that is well behaved.}  are strong indications that this proposal is well-defined and works for general $L$.  Moreover, the one case where we needed a regularization seems to be closely related to gauge transformations. Since the commutators involving the $\dot{\gen{S}}$ only generate $3 \rightarrow 2$ gauge interactions, this also suggests that there is no further problem with finiteness. However, while there clearly are no divergences from integrations over large $\lambda$, since the delta function of momentum has support on a compact region, we have not rigorously ruled out divergent contributions from integrating over momenta where the amplitude has singularities.  Especially since the $\text{N}^k$MHV amplitudes do take a more complicated form \cite{Drummond:2008cr} with additional collinear and multiparticle singularities, it would be worthwhile to at least compute $\gen{D}^{[1]}_{4 \rightarrow 2}$ which corresponds to the 6-particle NMHV amplitude and should be tractable. Even assuming finiteness, in principle it is also possible that the proposal maps polynomials to expressions including negative powers of $\lambda$ too. This seems unlikely because such interactions would have to be $\alg{psu}(2,2|4)$-invariant. Perhaps some more thought about constraints from superconformal symmetry would rule out this possibility.

\section{Conclusions \label{sec:conclusion}} 
This work has initiated the study of the complete planar $\mathcal{N}=4$ SYM dilatation generator spin chain representation beyond the one-loop dilatation generator. We have used superconformal symmetry to derive equations (\ref{eq:kidentity}-\ref{eq:parityandconjugation}) that give  $\gen{D}_{L \rightarrow 2}$ in terms of $\gen{D}_{L' \rightarrow 2}$  for $L' < L$. This shows that there is a unique solution for the leading $\gen{D}_{L \rightarrow 2}$ consistent with superconformal symmetry. The leading $\gen{D}_{2 \rightarrow L}$ follow from Hermitian conjugation. Recognizing that the symmetry constraints satisfied by  tree-level scattering amplitudes and the $\gen{D}_{L \rightarrow 2}$ are very similar, we proposed that  $\gen{D}_{L \rightarrow 2}$ is given by the scattering amplitude operator.  In addition to explicit verification in the case of $L=2,3$, with a few assumptions we proved that this proposal satisfies the superconformal symmetry constraints. However, we are lacking a proof that the amplitude operator maps polynomials to (finite) polynomials for larger $L$ as well. Even just a direct evaluation of the six-particle NMHV amplitude contributions ($\gen{D}_{4 \rightarrow 2}^{[1]}$) could be  useful confirmation that there are no subtleties for non-MHV contributions.

While we worked exclusively in the planar limit, the construction of Section \ref{sec:theconstruction} for $\gen{D}_{L \rightarrow 2}$ does not seem to rely fundamentally on the planar limit. Furthermore, the deformed superconformal symmetry representation of \cite{Bargheer:2009qu} for scattering amplitudes applies even for finite-rank gauge groups.  Therefore, the planar scattering amplitude-dilatation generator relation may extend to the complete finite-$N$ theory, which  includes all nonplanar contributions. Such a relation would relate the nonplanar $\gen{D}_{L \rightarrow 2}$ interactions to the complete tree-level amplitude $A_{L+2}$, which involves a weighted summation over color-ordered amplitudes with permuted arguments.

Similarly, since (maximal) supersymmetry was not central to the construction of Section \ref{sec:theconstruction}, we expect that the construction can be used for other superconformal gauge theories. In particular, it would be interesting to consider $\mathcal{N}=2$ superconformal QCD. Recently, \cite{Liendo:2011xb}  derived that theory's complete planar one-loop dilatation generator, showing that it is fixed by superconformal symmetry and can be written in terms of the harmonic action. It would therefore be good to check whether $\mathcal{N}=2$ superconformal QCD's dilatation generator also has length-changing interactions that are fixed by superconformal symmetry, and whether a (regularized) scattering operator gives the leading dilatation generator interactions for this theory.

Returning to planar $\mathcal{N}=4$ SYM, a major open problem is to compute the remaining $\gen{D}_{L \rightarrow L'}$. The next simplest generator is the ``two-loop'' dilatation generator $\gen{D}_{3 \rightarrow 3}$, which satisfies
\<
0 \eq \comm{\gen{S}_{1 \rightarrow 1}}{\gen{D}_{3 \rightarrow 3}} + \comm{\gen{S}_{2 \rightarrow 1}}{\gen{D}_{2 \rightarrow 3}}+ \comm{\gen{S}_{2 \rightarrow 2}}{\gen{D}_{2 \rightarrow 2}}, \nln
0 \eq \comm{\gen{Q}_{1 \rightarrow 1}}{\gen{D}_{3 \rightarrow 3}} + \comm{\gen{Q}_{1 \rightarrow 2}}{\gen{D}_{3 \rightarrow 2}}+ \comm{\gen{Q}_{2 \rightarrow 2}}{\gen{D}_{2 \rightarrow 2}},
\>
and similar equations with the dotted supersymmetry generators. In addition to the $\gen{D}_{2 \rightarrow 3}$ and $\gen{D}_{3 \rightarrow 2}$  obtained in this work, these constraints also involve  the currently unknown $2 \rightarrow 2$ supercharge interactions.  Moreover, because corrections to both $\gen{S}$ and $\gen{Q}$  appear here, it seems that $\gen{D}_{3 \rightarrow 3}$ does not equal (the integral of) six-particle scattering amplitudes\footnote{One could try doing such an integral to obtain  $3 \rightarrow 3$ interactions and then adding the Hermitian conjugate interactions. However, we see no reason that the dilatation generator can be written in this form, and even just in the $\alg{su}(2|3)$ sector the results of \cite{Beisert:2003ys} appear to be incompatible with this possibility.} since such an expression would automatically commute with $\gen{Q}_{1 \rightarrow 1}$.  Nonetheless, the connection to scattering amplitudes could still be important beyond leading order. For instance, given that tree-level amplitudes satisfy the BCFW recursion relation, it is reasonable to expect that the momentum localization integral and the translation  (back) to oscillator language gives ``dual BCFW'' recursion relations for $\gen{D}_{L \rightarrow 2}$. Then, in  oscillator language it would be trivial to obtain the Hermitian conjugate recursion relations for the $\gen{D}_{2 \rightarrow L}$, and possibly these two (conjectured) recursion relations could be combined to give relations valid for $\gen{D}_{3 \rightarrow 3}$ and general $\gen{D}_{L \rightarrow L'}$. The iterative structure in the $\alg{psu}(1,1|2)$ sector to three loops \cite{Zwiebel:2008gr} resembles such a picture. 

When checking superconformal symmetry, we encountered gauge interactions   $\gen{k}$,  $\gen{s}$, and $\dot{\gen{s}}$. It would be nice to identify the algebraic structure that encompasses the ordinary superconformal generators and these gauge interactions. One possibility is that these gauge interactions are related to the regularization of bilocal Yangian generators.  

Since (tree-level) amplitudes have Yangian symmetry, it should be straightforward to show that the $\gen{D}_{L \rightarrow 2}$ are consistent with Yangian symmetry. Significantly, this would enable a confirmation that the extra level-one Yangian generator $\hat{\gen{B}}$ of the spin chain $S$-matrix \cite{Matsumoto:2007rh,Beisert:2007ty} and of scattering amplitudes \cite{Beisert:2011pn} is also a symmetry of the dilatation generator.

More generally, given the multiple perspectives on scattering amplitudes that have been developed in recent years, the results of this work encourage seeking additional connections between scattering amplitudes  and the spectral problem. It would be particularly interesting if there were an analogous precise geometric relation at strong coupling between scattering amplitude and string energy calculations.

\subsection*{Acknowledgments}
I thank Niklas Beisert for helpful discussions and for sharing his observation of the connection between the tree-level four-particle MHV amplitude and the rotating oscillator form of the one-loop dilatation generator.  I also thank Ofer Aharony, John Schwarz and
Matthias Staudacher for useful discussions and correspondence. The research of the author was supported by a Lee A. DuBridge Postdoctoral Fellowship of the California Institute of Technology.  In addition, this work is supported in part by the DOE grant DE-FG03-92-ER40701.
\appendix

\section{The generalized gauge transformations \label{section:appendixgaugetransformations}}

This appendix describes how to compute the generalized gauge transformations $\gen{k}$, $\gen{s}$, and $\dot{\gen{s}}$, and gives analytic expressions for these gauge transformation generators. Afterwards, the  expansion of the commutator between $\gen{K}$ and $\delta \gen{D}$ in terms of local interactions appears.

To enable an analytic calculation of the gauge transformations we start by restricting to a $\alg{psu}(2,1|3)$ sector \cite{Beisert:2003jj}. The states of this sector have no   $\mathbf{d}^{\dagger 1}$ or $\mathbf{b}^{^\dagger 2}$ excitations. These conditions on the oscillators imply that in the classical linear approximation two Hermitian conjugate supercharges  $\dot{\gen{Q}}_1^2$ and $\dot{\gen{S}}_2^1$ annihilate the states of this sector. Equivalently, this sector satisfies  a $1/16$th BPS condition at $g=0$, and therefore this sector is closed to all orders in perturbation theory.  Then, in this sector, we have
\[
\gen{D}^{[1]}_{3 \rightarrow 2} = 2 \acomm{(\dot{\gen{Q}}_1^2)_{2 \rightarrow 2}}{(\dot{\gen{S}}_2^1)_{2 \rightarrow 1}}. \label{eq:d3inpsu213}
\]
Due to this relation, finding $(\dot{\gen{Q}}_1^2)_{2 \rightarrow 2}$ in this sector is the main step for obtaining analytic expressions for the gauge transformations.  $(\dot{\gen{Q}}_1^2)_{2 \rightarrow 2}$ is uniquely fixed by the $\alg{psu}(2,1|3) \times \alg{psu}(1|1)$ symmetry of this sector, and can be written  relatively compactly  using one new function:
\< \lefteqn{w(m_1, m_2;n_1, n_2) =} \\
& &   \sum_{i=0}^{m_1} \sum_{j=0}^{m_2} (-1)^{i+j}\frac{(m_1+n_1+1)!\,(m_2+n_2+1)!}{(m_1-i)!\, i! \, (m_2-j)!\, j!\, n_1!\, n_2!} \,  \frac{(n_1+i+1)^{-1}-(n_2+j+1)^{-1}}{n_1+n_2+i+j+2}.\notag
\>
Then we have
\begin{eqnarray}
 \lefteqn{\langle \vec{n}_1 \vec{n}_2|(\dot{\gen{Q}}^2_1)_{2 \rightarrow 2}|\vec{m}_1 \vec{m}_2 \rangle =  \delta(\vec{m}_1+\vec{m}_2- \vec{n}_1-\vec{n}_2+ \vec{\delta}_{12678} - \vec{\delta}_3) (-1)^{\sigma_{mn}} }
 \notag \\
& & \times \frac{m_{1,1}!\,m_{1,2}!\,m_{2,1}!\,m_{2,2}!}{n_{1,1}!\,n_{1,2}!\,n_{2,1}!\,n_{2,2}!}  \bigg(\theta(n_{1,1}-m_{1,1}-1)  -\theta(n_{1,2}-m_{1,2}-1)
\notag \\
 & &  + w(n_{1,1}-m_{1,1}-1,n_{1,2}-m_{1,2}-1;m_{1,1}, m_{1,2})
\notag \\
& &  -  w(n_{2,1}-m_{2,1}-1,n_{2,2}-m_{2,2}-1;m_{2,1}, m_{2,2}) +  w(m_{1,1},m_{1,2};m_{2,1},m_{2,2}) 
  \bigg),
\notag \\
\lefteqn{\sigma_{mn} =  n_{1,6}+ n_{1,8} +n_{2,7} + \sum_{k=6}^7\sum_{l=k+1}^8 (m_{2,k}m_{1,l} + n_{2,k}n_{1,l}) .}
 \end{eqnarray}
Here $\theta$ is the unit step function, with $\theta(0)=1$. Now (\ref{eq:d3inpsu213})   gives $\gen{D}^{[1]}_{3 \rightarrow 2}$, which we use to  compute analytically, for instance,  $(\gen{s}_{12})_{3 \rightarrow 2}$ within this sector. Then  Lorentz symmetry and R-symmetry uniquely fix  $(\gen{s}_{12})_{3 \rightarrow 2}$  for arbitrary $\alg{psu}(2,2|4)$ states. At this point, $(\gen{s}_{12})_{3 \rightarrow 2}$ lifts straightforwardly to  all $(\gen{s}_{\alpha C})_{3 \rightarrow 2}$. After simplifying we obtain 
\begin{eqnarray}
\lefteqn {\langle \vec{n}_1 \vec{n}_2 |(\gen{s}_{\alpha C})_{3 \rightarrow 2}|\vec{m}_1\vec{m}_2\vec{m}_3 \rangle =} 
\\ 
& & \delta(  \vec{m}_3-\vec{n}_2) s_{\alpha C}(\vec{m}_1,\vec{m}_2; \vec{n_1}) - (-1)^{m_{1,f}} \delta(  \vec{m}_1-\vec{n}_1) s_{\alpha C}(\vec{m}_2,\vec{m}_3; \vec{n_2}),
\notag \\
\lefteqn{s_{\alpha C}(\vec{m}_1,\vec{m}_2; \vec{n}) =\delta(\vec{m}_1+\vec{m}_2-\vec{n}-\vec{\delta}_{12}+\vec{\delta}_\alpha-\vec{\delta}_C)} \notag \\
& & \times \bigg( \Big(S_1(m_{2,1}+m_{2,2})-S_1(m_{1,1}+m_{1,2})\Big) \langle \vec{n} \, \vec{d} | \varepsilon_{\alpha \beta} \mathbf{a}_1^{\dagger \beta} \mathbf{d}_{2,C} \mathcal{H}^{(\half, \half)}|\vec{m}_1\vec{m}_2\rangle
\notag \\
& & + \frac{ 1}{n_1+n_2+1}  \langle \vec{n} \, \vec{d} | \varepsilon_{\alpha \beta} \mathbf{a}_1^{\dagger \beta} \mathbf{d}_{1,C} \mathcal{H}^{(0,0)}|\vec{m}_1\vec{m}_2\rangle \bigg),
\notag \\
\lefteqn{\vec{d} = (0,0,0,0,0,0,0,0).} \notag
\end{eqnarray}
$\mathcal{H}$ was defined in (\ref{eq:defineharmonicaction}), and again $S_1(n)$ is the $n$th ordinary harmonic number.
 By conjugation symmetry (\ref{eq:conjugation}) we also have
\begin{eqnarray}
\lefteqn {\langle \vec{n}_1 \vec{n}_2 |(\dot{\gen{s}}_{\dot{\alpha}}^C)_{3 \rightarrow 2}|\vec{m}_1\vec{m}_2\vec{m}_3 \rangle =} 
\\ 
& & \delta(  \vec{m}_3-\vec{n}_2) \dot{s}_{\dot{\alpha}}^C(\vec{m}_1,\vec{m}_2; \vec{n_1}) - (-1)^{m_{1,f}} \delta(  \vec{m}_1-\vec{n}_1) \dot{s}_{\dot{\alpha}}^C(\vec{m}_2,\vec{m}_3; \vec{n_2}),
\notag \\
\lefteqn{\dot{s}_{\dot{\alpha}}^C(\vec{m}_1,\vec{m}_2; \vec{n}) =\delta(\vec{m}_1+\vec{m}_2-\vec{n}-\vec{\delta}_{34}+\vec{\delta}_{\dot{\alpha}}-\vec{\delta}_{5678} +\vec{\delta}_C)} \notag \\
& & \times \bigg( \Big(S_1(m_{2,3}+m_{2,4})-S_1(m_{1,3}+m_{1,4})\Big) \langle \vec{n} \, \vec{c} | \varepsilon_{\dot{\alpha}\dot{ \beta}} \mathbf{b}_1^{\dagger \dot{\beta}} \mathbf{d}^{\dagger C}_2 \mathcal{H}^{(\half, \half)}|\vec{m}_1\vec{m}_2\rangle
\notag \\
& & + \frac{ 1}{n_3+n_4+1}  \langle \vec{n} \, \vec{c} | \varepsilon_{\dot{\alpha}\dot{ \beta}} \mathbf{b}_1^{\dagger \dot{\beta}} \mathbf{d}^{\dagger C}_1  \mathcal{H}^{(0,0)}|\vec{m}_1\vec{m}_2\rangle \bigg),
\notag \\
\lefteqn{\vec{c} = (0,0,0,0,1,1,1,1).} \notag
\end{eqnarray}
The final gauge transformations, $\gen{k}_{\alpha \dot{\beta}}$ then follow from (\ref{eq:kgaugefromcommutator}),
\<
 (\gen{k}_{\alpha\dot{\beta}})_{3 \rightarrow 2} \eq \acomm{(\gen{s}_{\alpha 1})_{3 \rightarrow 2}}{(\dot{\gen{S}}_{\dot{\beta}}^1)_{1\rightarrow 1}}+ \acomm{(\gen{S}_{\alpha1})_{1\rightarrow 1}}{(\dot{\gen{s}}_{\dot{\beta}}^1)_{3 \rightarrow 2}} ,
\nln
(\gen{k}_{\alpha\dot{\beta}})_{4 \rightarrow 2} \eq \acomm{(\gen{s}_{\alpha 1})_{3 \rightarrow 2}}{(\dot{\gen{S}}_{\dot{\beta}}^1)_{2\rightarrow 1}}+ \acomm{(\gen{S}_{\alpha1})_{2\rightarrow 1}}{(\dot{\gen{s}}_{\dot{\beta}}^1)_{3 \rightarrow 2}} .
 \>

Finally, we include the complete expansion of the commutator of $\gen{K}$ and $\delta \gen{D}$ (\ref{eq:briefkcommutator}),
\<  \label{eq:fullkcommutator}
\lefteqn{\comm{\gen{K}}{g^{-2}\, \delta\gen{D}}_{L \rightarrow 2} =  \delta_{L3} \gen{k}_{3 \rightarrow 2} +  \delta_{L4} \gen{k}_{4 \rightarrow 2} } \\
\eq \comm{\gen{K}_{1 \rightarrow 1}}{\gen{D}_{L \rightarrow 2}} + \delta_{L\neq3} \, \comm{\gen{K}_{2 \rightarrow 1}}{\gen{D}_{(L-1) \rightarrow 2}} + \theta(L-4) \, \comm{\gen{K}_{3 \rightarrow 1}}{\gen{D}_{(L-2) \rightarrow 2}} \notag \\
\eq   \sum_{i=1}^2 \gen{K}_{1 \rightarrow 1}(i) \gen{D}_{L \rightarrow 2}(1, 2, \ldots L)-  \sum_{i=1}^L \gen{D}_{L \rightarrow 2}(1, 2, \ldots L) \gen{K}_{1 \rightarrow 1}(i) \notag \\
& + &  \delta_{L\neq2}   \bigg( \gen{K}_{2 \rightarrow 1}(2,3) \gen{D}_{(L-1) \rightarrow 2}(1, \ldots L-1) + \gen{K}_{2 \rightarrow 1}(1,2) \gen{D}_{(L-1) \rightarrow 2}(2,  \ldots L)  \notag \\
&  & \quad - \sum_{i=1}^{L-1} \gen{D}_{(L-1) \rightarrow 2}(1, 2, \ldots (L-1)) \gen{K}_{2 \rightarrow 1}(i,i+1) \bigg) \notag \\
& + &  \delta_{L>3}   \bigg( \gen{K}_{3 \rightarrow 1}(2,3,4) \gen{D}_{(L-2) \rightarrow 2}(1, \ldots L-2) + \gen{K}_{3 \rightarrow 1}(1,2,3) \gen{D}_{(L-2) \rightarrow 2}(3,  \ldots L)  \notag \\
&  & \quad - \sum_{i=1}^{L-2} \gen{D}_{(L-2) \rightarrow 2}(1, 2, \ldots (L-2)) \gen{K}_{2 \rightarrow 1}(i,i+1,i+2) \bigg) \notag \\ 
& + & \half \delta_{L3} \bigg( \gen{K}_{2 \rightarrow 1}(1,2) \gen{D}_{2\rightarrow 2}(1, 2) +  \gen{K}_{2 \rightarrow 1}(2,3) \gen{D}_{2\rightarrow 2}(2,3)\bigg) \notag \\
& + & \half \delta_{L4} \bigg( \gen{K}_{3 \rightarrow 1}(1,2,3) (\gen{D}_{2\rightarrow 2}(1, 2)+ \gen{D}_{2\rightarrow 2}(2,3))  \notag \\
& & \quad +  \gen{K}_{3 \rightarrow 1}(2,3,4) (\gen{D}_{2\rightarrow 2}(2, 3)+ \gen{D}_{2\rightarrow 2}(3,4)) \bigg). \notag
\>
%


\end{document}